\input psfig.sty
\def\Box{\kern1pt\vbox{\hrule height 1.2pt\hbox{\vrule width 1.2pt\hskip 3pt
   \vbox{\vskip 6pt}\hskip 3pt\vrule width 0.6pt}\hrule height 0.6pt}\kern1pt}
\def\tp{t^{\prime}}
\def\tpp{t^{\prime\prime}}
\def\gtwid{\mathrel{\raise.3ex\hbox{$>$\kern-.75em\lower1ex\hbox{$\sim$}}}}
\def\ltwid{\mathrel{\raise.3ex\hbox{$<$\kern-.75em\lower1ex\hbox{$\sim$}}}}
\documentclass[12pt]{article}
\input{psfig.sty}
\begin{document}
\begin{titlepage}
\begin{flushright}
hep-ph/9712331 \\ UFIFT-HEP-97-5 \\ CRETE-97-15
\end{flushright}
\vspace{.4cm}
\begin{center}
\textbf{Non-Perturbative Models For \\ 
The Quantum Gravitational Back-Reaction On Inflation}
\end{center}
\begin{center}
N. C. Tsamis$^{\dagger}$
\end{center}
\begin{center}
\textit{Department of Physics \\ University of Crete \\ Heraklion, GR-71003
GREECE}
\end{center}
\begin{center}
R. P. Woodard$^*$
\end{center}
\begin{center}
\textit{Department of Physics \\ University of Florida \\ 
Gainesville, FL 32611 USA}
\end{center}
\begin{center}
ABSTRACT
\end{center}
\hspace*{.5cm}
We consider a universe in which inflation commences because of a positive 
cosmological constant, the effect of which is progressively screened by the
interaction between virtual gravitons that become trapped in the expansion of
spacetime. Perturbative calculations have shown that screening becomes 
non-perturbatively large at late times. In this paper we consider effective 
field equations which can be evolved numerically to provide a non-perturbative 
description of the process. The induced stress tensor is that of an effective
scalar field which is a non-local functional of the metric. We use the known
perturbative result, constrained by general principles and guided by a physical
description of the screening mechanism, to formulate a class of ans\"{a}tze for
this functional. A scheme is given for numerically evolving the field equations
which result from a simple ans\"{a}tz, from the beginning of inflation past the
time when it ends. We find that inflation comes to a sudden end, producing a 
system whose equation of state rapidly approaches that of radiation. Explicit
numerical results are presented.
\begin{flushleft}
PACS numbers: 04.60.-m, 98.80.Cq
\end{flushleft}
\vspace{.4cm}
\begin{flushleft}
$^{\dagger}$ e-mail: tsamis@physics.uch.gr \\
$^*$ e-mail: woodard@phys.ufl.edu
\end{flushleft}
\end{titlepage}

\section{Introduction}

Perturbation theory is an immensely gratifying tool. It almost always provides
quantitative answers for how a known system changes with the inclusion of a
small, new effect. The great frustration in using the technique is that its 
answers become unreliable precisely when they are most interesting: when the 
new effect causes major changes. Our recent study of the quantum gravitational
back-reaction on inflation \cite{tw1} illustrates both the utility of 
perturbation theory, and the frustration of not being able to push it further.

The unperturbed system in our case is classical general relativity, the
Lagrangian for which is:
\begin{equation}
{\cal L} = {1 \over 16 \pi G} \left(R - 2 \Lambda\right) \sqrt{-g} \; .
\end{equation}
Here $G$ is Newton's constant and $\Lambda$ is the cosmological constant, 
assumed positive. On a spatially flat manifold the invariant element for a 
homogeneous and isotropic universe can be written in co-moving coordinates:
\begin{equation}
{\widehat g}_{\mu\nu}(t,{\vec x}) dx^{\mu} dx^{\nu} = - dt^2 + e^{2 b(t)}
d{\vec x} \cdot d{\vec x} \; . \label{eq:element}
\end{equation}
And the classical solution is:
\begin{equation}
b_{\rm class}(t) = H t \; , \label{eq:classical}
\end{equation} 
where $H \equiv (\Lambda/3)^{1/2}$ is the Hubble constant. If we specialize 
to the manifold $T^3 \times \Re$, where each of the coordinate radii is 
$H^{-1}$, then the 3-volume:
\begin{equation}
V(t) = H^{-3} e^{3 H t} \; ,
\end{equation}
is finite but grows exponentially.

The perturbation we seek to study is the gravitational interaction between
virtual infrared gravitons that become trapped in the expansion of spacetime 
and get pulled apart. Although we have computed this exactly at the lowest 
non-trivial order in perturbation theory \cite{tw1,tw2}, an intuitive 
understanding of the effect is necessary if we are to abstract it beyond the 
perturbative regime. The physical picture is that virtual gravitons of 
sufficiently long physical wave length are torn apart by inflation. This is the
phenomenon of {\it superadiabatic amplification}, first studied by Grishchuk in
1974 \cite{grish1}. Although infrared gravitons are continually produced in 
this way, the volume of space expands so rapidly that the energy density of 
these gravitons remains a constant --- and rather small --- fraction of 
$\Lambda/(8\pi G)$.\footnote{It is easy to show that there is on average one 
infrared graviton per Hubble volume. Even for inflation on the GUT scale this
is only about $10^{-11}$ of the energy density of the cosmological constant.} 
However, as each graviton pair recedes, the intervening space is filled by 
their long range gravitational potentials. These potentials persist even after 
the gravitons that engendered them have reached cosmological separations. As 
new pairs are ripped apart, their potentials add to those already present. This
is a secular effect and it obviously continues as long as inflation does. 
Because gravity is attractive the effect tends to counteract inflation --- and 
hence to screen the cosmological constant. 

One might expect similar results from quanta other than gravitons but this
is not so. To experience superadiabatic amplification a particle must be 
effectively massless with respect to $H$, and it must not possess classical
conformal invariance \cite{grish1,grish2}.\footnote{Massive particles are short 
range, so their virtual quanta seldom get far enough apart to become trapped 
in the expansion of spacetime. Conformally invariant particles are incapable, 
locally, of distinguishing between the conformally flat classical background 
(\ref{eq:classical}) and flat space.} One or the other of these two conditions
excludes every other known particle and most of the conjectured ones. The
only contender, besides the graviton, is a massless, minimally coupled scalar. 
These do experience superadiabatic amplification, but global conformal 
invariance prevents them from inducing a gravitational interaction comparable 
to that of gravitons \cite{tw5}. One might get a strong effect if such a 
scalar had non-derivative self-interactions, but it is difficult to understand 
why these would not also induce a substantial mass.

Screening affords a simple and satisfying reformulation of inflationary
cosmology and a beautiful resolution to the associated problems of fine
tuning. Inflation starts, in this scheme, because the cosmological constant 
is positive and not unreasonably small. Inflation eventually ends due to the 
self-gravitation of virtual gravitons which have become trapped in the 
superluminal expansion of spacetime. Inflation lasts for a very long time 
because gravitational interactions are weak, even at the GUT scale. One can 
be indifferent about adding matter because gravitons are the unique 
phenomenologically viable quanta which induce screening. The only thing to 
{\it avoid}, in this scheme, is introducing an inflaton field and fine tuning 
its potential! Best of all, the infrared character of the screening mechanism 
means that it can be studied reliably using quantum general relativity, in 
spite of the theory's lack of perturbative renormalizability and without 
regard to what happens at the Planck scale.\footnote{Infrared phenomena can 
always be studied using the low energy effective theory. This is why Bloch and 
Nordsieck \cite{bn} were able to resolve the infrared problem in QED before 
the theory's renormalizability was suspected. It is also why Weinberg 
\cite{wein} was able to give a similar resolution for the infrared problem of 
quantum general relativity with zero cosmological constant. And it is why 
Feinberg and Sucher \cite{fs} were able to compute the long range force 
induced by neutrino exchange using Fermi Theory.}

But there is a problem: the quantum gravitational back-reaction can only be 
studied perturbatively so long as it is weak. This regime is not without 
interest. For example, perturbative analysis shows that inflation lasts a 
long time \cite{tw4} and that conventional matter is incapable of competing 
with the quantum gravitational back-reaction \cite{tw5}. If one assumes a 
sudden end to inflation --- which is certainly supported by the perturbative 
results \cite{tw4} --- then it should be possible to predict the spectrum of 
density fluctuations in the perturbative regime. However, the most interesting 
questions lie frustratingly beyond the point where perturbation theory is 
valid. Hence the need for a non-perturbative model.

Our strategy for creating such a model is to infer the induced stress tensor:
\begin{equation}
T_{\mu\nu}[g] = {1 \over 8 \pi G} \left(R_{\mu\nu} - \frac12 g_{\mu\nu} R +
g_{\mu\nu} \Lambda \right) \; , \label{eq:induced}
\end{equation}
as a non-local functional of the metric which correctly reproduces the known
perturbative effect and which captures the physical origin of screening 
generally. Of course there is some ambiguity in this, but surprisingly little
of any significance. Given an ans\"atz we can numerically integrate the field 
equations as far into the future as is desired. What we find, for a simple 
ans\"atz, is that inflation ends suddenly over a period of about five 
e-foldings, following which the equation of state rapidly approaches that of 
pure radiation.

This paper consists of eight sections, of which the first is drawing to a 
close.  In Section 2 we show that, for the purposes of cosmology, the induced 
stress tensor can be parameterized as that of an effective scalar field which 
is a non-local functional of the metric. We also derive what this functional 
must be during the perturbative regime. In Section 3 we enumerate six 
principles which constrain the scalar functional generally. Section 4 gives a 
semi-quantitative model of screening which is of course the ultimate physical 
motivation for the choice of scalar. We discuss various ans\"atze for the 
scalar in Section 5. Section 6 describes our scheme for numerically integrating
the dynamical system resulting from a simple ans\"atz. We also report explicit
results. In Section 7 we reconstruct the potential of the effective scalar, 
analytically for large or small values of the scalar and numerically for any
value. Section 8 is a discussion of our results.

\section{Effective scalar stress tensor}

The point of this section is to show that, for the purposes of cosmology,
we can model the induced stress tensor ({\ref{eq:induced}) as that of a scalar 
field which is itself a non-local functional of the metric:
\begin{equation}
T_{\mu\nu}[g] = \partial_{\mu} \phi[g] \partial_{\nu} \phi[g] - g_{\mu\nu}
\left(\frac12 g^{\rho\sigma} \partial_{\rho} \phi[g] \partial_{\sigma} \phi[g]
+ V(\phi[g])\right) \; . \label{eq:scalarT}
\end{equation}
We will also show that, since the potential can always be chosen to enforce
conservation, one really needs only the functional $\phi[g]$. And we will
use the known perturbative results \cite{tw4} to derive what $\phi[g]$ must
be when specialized to the classical background (\ref{eq:classical}).

There are three senses in which one might discuss the induced stress tensor,
or any other functional of the metric. The first sense is generally for an 
arbitrary metric; the second is as a functional of $b(t)$ for a spatially flat,
homogeneous and isotropic metric; and the third is as an explicit function of 
time for the case of perturbative corrections. When the basic symbol appears 
unadorned we mean the general quantity; the presence of a hat implies 
specialization to the spatially flat, homogeneous and isotropic form; and the 
perturbative version of the same quantity is denoted by a tilde. For example, a
general invariant element is represented thus:
\begin{equation}
ds^2 = g_{\mu\nu}(t,{\vec x}) dx^{\mu} dx^{\nu} \; .
\end{equation}
Specializing to flat, homogeneous and isotropic spacetimes gives:
\begin{equation}
d{\widehat s}^2 = -dt^2 + \exp[2 b(t)] d{\vec x} \cdot d{\vec x} \; .
\label{eq:secondel}
\end{equation}
And we express the perturbative result as follows \cite{tw4}:
\begin{equation}
d{\widetilde s}^2 = -dt^2 +  \sqrt{1 + A(t)} \; \exp\left[2 H t \left\{1 + 
D(t)\right\}\right] \; , \label{eq:firstel}
\end{equation}
where perturbative expansions for the functions $D(t)$ and $A(t)$ are:
\begin{eqnarray}
D(t) & = & +{19 \over 2} (\epsilon H t)^2 + O\left((\epsilon H t)^3\right) 
\; , \label {eq:D} \\
A(t) & = & -{172 \over 9} \epsilon^2 (H t)^3 + O\left(\epsilon^3 (H t)^4\right)
\; , \label{eq:A}
\end{eqnarray}
The small parameter in these expansions is $\epsilon \equiv G \Lambda/{3 \pi}$.
We will assume that it can be as big as $10^{-12}$ or as small as $10^{-68}$.

Although deep intuition about the induced stress tensor derives from its 
dependence upon a general metric, we cannot hope to say much about the 
micro-structure of quantum gravity. Nor is this necessary. Screening is a 
phenomenon of cosmological scales, so we need only a model that is accurate for
spatially flat, homogeneous and isotropic metrics ({\ref{eq:secondel}). The
isometries of these metrics imply that only ${\widehat T}_{00}[b]$ and 
${\widehat T}_{ij}[b]$ can be non-zero, and that both are functions of time 
alone. We shall parameterize them in the usual way as an induced energy density
$\rho(t)$ and an induced pressure $p(t)$:
\begin{equation}
{\widehat T}_{00}[b](t) \equiv \rho(t) \qquad , \qquad {\widehat T}_{ij}[b](t)
\equiv g_{ij} p(t) \; .
\end{equation}
We first show that $\rho(t)$ and $p(t)$ can be chosen so as to support any
evolution for $b(t)$. Then we show that it is always possible to choose the
scalar field $\phi[g]$ and its potential $V(\phi)$ to give the desired energy 
density and pressure.

The non-trivial components of the effective field equations are:
\begin{eqnarray}
3 \dot{b}^2 & = & 3 H^2 + 8 \pi G \rho \; , \label{eq:rhotot} \\
-2 \ddot{b} -3 \dot{b}^2 & = & -3 H^2 + 8 \pi G p \; . \label{eq:ptot}
\end{eqnarray}
Although one usually regards these as equations for $b(t)$ in terms of 
$\rho(t)$ and $p(t)$, we can take the converse view:
\begin{eqnarray}
\rho(t) & = & {1 \over 8 \pi G} \left(3 \dot{b}^2(t) - 3 H^2\right) \; , 
\label{eq:rhob} \\
p(t) & = & {1 \over 8 \pi G} \left(-2 \ddot{b}(t) - 3 \dot{b}^2(t) + 3 H^2
\right) \; . \label{eq:pressureb}
\end{eqnarray}
The physical import of these equations is that one can find $\rho(t)$ and 
$p(t)$ so as to support {\it any} evolution for $b(t)$.

In a homogeneous and isotropic background, the energy density and pressure of a 
scalar ${\widehat \phi}[b]$ are:
\begin{eqnarray}
\rho & = & {1 \over 2} \left({d{\widehat \phi} \over dt}\right)^2 + 
V({\widehat \phi}) \; , \\
p & = & {1 \over 2} \left({d{\widehat \phi} \over dt}\right)^2 - 
V({\widehat \phi}) \; .
\end{eqnarray}
Combining this with ({\ref{eq:rhob}-\ref{eq:pressureb}) we see find that an
arbitrary evolution $b(t)$ can be supported by making the following choices
for the scalar and its potential:
\begin{eqnarray}
\left({d{\widehat \phi} \over dt}\right)^2 & = & {1 \over 8 \pi G} \left( -2 
\ddot{b}\right) \; , \label{eq:phidef} \\
V & = & {1 \over 8 \pi G} \left(\ddot{b} + 3 \dot{b}^2 - 3 H^2 \right) \; . 
\label{eq:Vdef}
\end{eqnarray}
Given an explicit function $b(t)$ one constructs $V(\phi)$ by solving
the differential equation (\ref{eq:phidef}) for the scalar as an explicit 
function of time, call it ${\widehat \phi}[b](t) = f(t)$. We then invert this 
relation to express time as a function of ${\widehat \phi}$, $t = 
f^{-1}({\widehat \phi})$. The potential $V(\phi)$ is found by evaluating
relation (\ref{eq:Vdef}) for $t = f^{-1}(\phi)$. 

The construction is completed by giving a functional of the metric which agrees
with ${\widehat \phi}(t)$ for the particular choice of $b(t)$. There are many
solutions. Perhaps the simplest is obtained from $P[g](t,{\vec x})$, the 
invariant volume of the past light cone of the point $(t,{\vec x})$. For a
spatially flat, homogeneous and isotropic metric, this is a monotonically
increasing function of the co-moving time $t$:
\begin{equation}
{\widehat P}[b](t) = \frac43 \pi \int_0^t d\tp e^{3 b(\tp)} \; \left( 
\int_{\tp}^t d\tpp e^{-b(\tpp)} \right)^3 \; ,
\end{equation}
so we can invert the relation. Suppose that for the specific function $b(t)$ we
get ${\widehat P}[b](t) = \pi(t)$. Then time is $t = \pi^{-1}({\widehat P})$, 
and the scalar for a general metric could be taken as:
\begin{equation}
\phi[g] = f\left(\pi^{-1}(P[g])\right) \; .
\end{equation}
Since all this can be done for any function $b(t)$, we lose nothing by assuming
that the induced stress tensor has the scalar form (\ref{eq:scalarT}).

In the preceding discussion we inverted the proper order of things to show 
that the scalar stress tensor (\ref{eq:scalarT}) can describe any homogeneous 
and isotropic geometry. That point having been made, we can return to the usual
dynamical problem of inferring $g_{\mu\nu}$ from $T_{\mu\nu}[g]$. For a 
homogeneous and isotropic universe the non-trivial equations are:
\begin{eqnarray}
3 \dot{b}^2 & = & 3 H^2 + 8 \pi G \left\{{1 \over 2} \left({d{\widehat \phi}[b]
\over dt}\right)^2 + V\left({\widehat \phi}\right)\right\} \; , 
\label{eq:Friedman} \\
-2 \ddot{b} - 3\dot{b}^2 & = & - 3 H^2 + 8 \pi G \left\{{1 \over 2} 
\left({d{\widehat \phi}[b] \over dt}\right)^2 - V\left({\widehat \phi}\right)
\right\} \; . \label{eq:Gij}
\end{eqnarray}
In dynamical terms, equation (\ref{eq:Friedman}) is actually a constraint. If 
it is true initially then time evolution and energy conservation conspire to 
keep it true. The dynamical equation for $b(t)$ could be taken to be 
(\ref{eq:Gij}), but it is more convenient to add (\ref{eq:Friedman}):
\begin{equation}
\ddot{b} = - 4 \pi G \left({d{\widehat \phi}[b] \over dt}\right)^2 \; .
\label{eq:dynamical}
\end{equation} 
Note that the scalar potential has dropped out. {\it It therefore follows that
a model is specified by giving the scalar} $\phi[g]$ {\it as a functional of 
the metric.} 

If the potential is desired it can be determined by the round-about process of
first solving  (\ref{eq:dynamical}) for $b(t)$ and substituting to find 
${\widehat \phi}[b](t)$. One then inverts to express $t$ as a function of 
${\widehat \phi}$, and finally substitutes into the constraint 
(\ref{eq:Friedman}):
\begin{equation}
V\left({\widehat \phi}\right) = -{1 \over 2} \left({d{\widehat \phi} \over dt}
\right)^2 + {3 \over 8 \pi G} \left(\dot{b}^2 - H^2\right) \; .
\end{equation}
This turns out to be much easier than it might seem. In Section 7 we will 
obtain analytic expressions for $V(\phi)$ in the perturbative regime and in the
regime of asymptotically late times. We will also carry out the process 
numerically over the full range of evolution.

It remains to work out ${\widetilde \phi}(t)$ during the perturbative regime. 
Using relation (\ref{eq:firstel}) we can express the second time derivative of 
$b(t)$ in terms of the functions $D(t)$ and $A(t)$:
\begin{equation}
\ddot{b}(t) = H \left[2 \dot{D}(t) + t \ddot{D}(t)\right] + {1 \over 2} \left(
{\ddot{A}(t) \over 1 + A(t)}\right) - {1 \over 2} \left({\dot{A}(t) \over 1 + 
A(t)}\right)^2 \; .
\end{equation}
It is easy to see from the perturbative expansions (\ref{eq:D}) and 
(\ref{eq:A}) that only the third term matters at any time during the 
perturbative regime \cite{tw4}. Comparing with expression (\ref{eq:dynamical}) 
we infer:
\begin{equation}
\left({d{\widetilde \phi} \over dt}\right)^2 \approx {1 \over 8 \pi G} \; 
\left({\dot{A} \over 1 + A}\right)^2 \; .
\end{equation}
Making an arbitrary choice of sign and using the fact that only the first term
in the expansion of $A(t)$ matters, we obtain the following formula for the
scalar during the perturbative regime:
\begin{eqnarray}
{\widetilde \phi}(t) & \approx & - {1 \over \sqrt{8 \pi G}} \; \ln\left[1 +
A(t)\right] \; , \\
& \approx & - {1 \over \sqrt{8 \pi G}} \; \ln\left[1 - {172 \over 9} \epsilon^2 
(H t)^3\right] \; , \label{eq:pertF}
\end{eqnarray}
where the small parameter is $\epsilon \equiv G \Lambda/(3 \pi)$.

\section{General principles}

In the previous section we saw that, on cosmological scales, the induced stress
tensor can be taken as that of a scalar field, $\phi[g]$ which is itself a
non-local functional of the metric. We saw further that it is really only 
necessary to give this functional, since the associated potential is determined
by conservation. In fact we actually require only the restriction ${\widehat
\phi}[b]$ of this functional to a flat, homogeneous and isotropic geometry.
However, powerful constraints exist on how the scalar can depend upon a general
metric. The purpose of this section is to state these constraints.

\vspace{.3cm}
\noindent {\it 1. Causality} \newline
\indent We are actually going to {\it guess} the induced stress tensor but, 
were we to {\it compute} it, $T_{\mu\nu}[g](x)$ would come from Schwinger's 
effective action for expectation values \cite{schw,rj}, not from the more 
common, ``in''-``out'' effective action. One consequence is that $T_{\mu\nu}[g]
(x)$ --- and hence also $\phi[g](x)$ --- can only depend upon the metric at 
points $y^{\mu}$ which lie within the past light cone of $x^{\mu}$. It is worth 
noting that the effective field equations for ``in''-``out'' matrix elements 
must be symmetric: if they depend upon a field at $x^{\mu} - {\Delta x}^{\mu}$ 
then they must also depend upon fields at $x^{\mu} + {\Delta x}^{\mu}$. This is 
avoided in Schwinger's method because his effective action really depends upon 
{\it two} fields, one the background during forward time evolution and the 
other the background during evolution back to the initial state. The effective 
field equations are obtained by varying with respect to (either) one of these 
fields and then setting them equal. This is what breaks the forward-backwards 
symmetry of the ``in''-``out'' field equations. Causality arises because the 
forward and backward evolutions interfere destructively outside the past light 
cone.

\vspace{.3cm}
\noindent {\it 2. General Coordinate Invariance} \newline
\indent Because the dynamics of quantum general relativity are general
coordinate invariant, non-invariance can only enter the effective field
equations from the gauge in which the initial state was specified. In other
words, $\phi[g]$ must be invariant up to surface terms. This means that
it must consist of a covariant local part plus non-local operators, such as 
the retarded propagator, acting on local functions of the Riemann tensor and 
its covariant derivatives. Although one can form many covariants from the
Riemann tensor and its derivatives, only a few are distinct in a homogeneous 
and isotropic universe. To see this, note first that the spacetime is 
conformally flat. This means the Weyl tensor vanishes and one can express 
the Riemann tensor in terms of the Ricci tensor:
\begin{equation}
{\widehat R}_{\rho\sigma\mu\nu} = \frac12 \left({\widehat g}_{\rho\mu} 
{\widehat R}_{\sigma\nu} - {\widehat g}_{\mu\sigma} {\widehat R}_{\nu\rho} + 
{\widehat g}_{\sigma\nu} {\widehat R}_{\rho\mu} - {\widehat g}_{\nu\rho} 
{\widehat R}_{\mu\sigma}\right) - \frac16 \left({\widehat g}_{\rho\mu} 
{\widehat g}_{\sigma\nu} - {\widehat g}_{\rho\nu} {\widehat g}_{\sigma
\mu}\right) {\widehat R} \; .
\end{equation}
Although we will express $\phi[g]$ as an invariant functional of a general
metric, we need not worry about the distinction between ans\"{a}tze which agree
for a spatially flat, homogeneous and isotropic universe.

\vspace{.3cm}
\noindent {\it 3. Stability of the initial value problem} \newline
\indent Since the quantum field theoretic problem was well posed given only 
the initial state wavefunctional, it must be that the associated effective
field equations can be evolved forward from $t=0$ knowing only the metric and 
its first time derivative. This limits the local terms in $T_{\mu\nu}[g]$ to 
those which contain at most second time derivatives of the metric. Since the
induced stress tensor contains derivatives of the effective scalar, the local
part of $\phi[g]$ can have at most first derivatives. Stability also imposes 
requirements on non-local terms which are differentiated or which can give 
local terms by partial integration.

\vspace{.3cm}
\noindent {\it 4. Non-locality} \newline
\indent A universe which is initially inflating will continue to inflate 
unless $T_{\mu\nu}[g]$ is a non-local functional of the metric. To see this,
assume the converse. Using the previous principle we can then constrain 
$T_{\mu\nu}[g]$ to consist of second rank functions of the Riemann tensor. 
Note that in a locally de Sitter geometry the Riemann tensor can be 
written in terms of the Ricci scalar. One way of expressing this is by
saying that the following tensor vanishes:
\begin{equation}
V_{\rho\sigma\mu\nu} \equiv R_{\rho\sigma\mu\nu} - \frac1{12} 
\left(g_{\rho\mu} g_{\sigma\nu} - g_{\rho\nu} g_{\sigma\mu}\right) R \; .
\end{equation}
This fact can be used to write any second rank tensor function of the 
Riemann tensor as a term which vanishes in a locally de Sitter geometry 
plus a function of the Ricci scalar times the metric. For example, consider 
the partially contracted product of two Riemann tensors:
\begin{equation}
R_{\mu}^{~\alpha\beta\gamma} R_{\nu\alpha\beta\gamma} = V_{\mu}^{~\alpha
\beta\gamma} V_{\nu\alpha\beta\gamma} + \frac13 R \: V^{\rho}_{~~\mu \rho \nu}
+ \frac1{16} R^2 g_{\mu\nu} \; .
\end{equation}
But the initial condition of our problem is a locally de Sitter geometry.
Therefore all terms of the first type vanish, and terms of the second type
simply renormalize the cosmological constant. We have already defined our 
cosmological constant to absorb any terms of the second type, so local 
effective field equations would have $R_{\mu \nu} = \Lambda g_{\mu\nu}$ as a 
solution for all time. Since we can actually see inflation begin to slow in 
perturbation theory it follows that non-local terms must be responsible. The 
same argument works later on, even after the effective Hubble constant has 
been substantially reduced: the effect would stop without non-local terms. 
So the important part of the induced stress tensor must be non-local.

\vspace{.3cm}
\noindent {\it 5. Dimensional Analysis} \newline
\indent The induced stress tensor has the dimensions of length$^{-4}$, so
$\phi[g]$ goes like length$^{-1}$. This seemingly trivial fact conceals a 
surprisingly powerful constraint. The most important quantities from which 
$\phi[g]$ can be constructed have the following dimensionalities:
\begin{eqnarray}
G \sim {\rm length}^2 & , & R_{\rho\sigma\mu\nu} \sim {\rm length}^{-2} \; ,
\nonumber \\
\Lambda \sim {\rm length}^{-2} & , & {1 \over \Box} \sim {\rm length}^2 \; ,
\end{eqnarray}
where $\Box$ is the scalar d'Alembertian:
\begin{equation}
\Box \equiv {1 \over \sqrt{-g}} \; \partial_{\mu} \left( g^{\mu\nu} \sqrt{-g}
\; \partial_{\nu}\right) \; .
\end{equation}
Note that the dimensionless quantity $G \Lambda$ is less than about $10^{-11}$,
even for GUT scale inflation. Further, the curvature is guaranteed to be of 
order $\Lambda$ during the perturbative regime, and it had better be 
considerably smaller at late times. This means that terms with too many powers 
of $G$ are likely to be negligible unless they accompany non-local growth 
factors such as $1/\Box$.

\vspace{.3cm}
\noindent {\it 6. The Flat Space Limit} \newline
\indent When $\Lambda = 0$ we know that the vacuum is stable, so the 
``in''-``out'' matrix element of an operator agrees with its expectation value.
We also know that ``in''-``out'' amplitudes with $\ell$ loop amplitudes contain
at most $\ell$ infrared logarithms \cite{wein}. This means that the most 
infrared singular term which can remain in the $\Lambda = 0$ ``in''-``out'' 
effective action has the form:
\begin{equation}
\Gamma_{\rm flat}[g] \sim \int d^4x \sqrt{-g} \: R \: \left[ \ln(\Box) 
\right]^{\ell} R \; .
\end{equation}
Terms whose field dependence is more singular must possess positive powers
of $\Lambda$, for example:
\begin{equation}
\Lambda \int d^4x \sqrt{-g} \: R \: {1 \over \Box} \: R \; .
\end{equation}
And we must of course avoid inverse powers of $\Lambda$.

\section{The physics of screening}

The most important constraint on the functional $\phi[g]$ is that it should
correctly reflect the physics of screening. Of course choosing the scalar so
that its perturbative restriction agrees with (\ref{eq:pertF}) automatically
enforces this during the perturbative regime, so any additional information 
must come from understanding the mechanism of screening for an arbitrary 
homogeneous and isotropic background. That is the purpose of this section. Our
procedure is to work first in the classical background, where results can be 
checked against perturbation theory, and then generalize. We begin by giving a
simple derivation of the phenomenon of superadiabatic amplification 
\cite{grish1,grish2}, whereby the 0-point energy of infrared graviton modes is 
vastly enhanced over the familiar $\frac12 \hbar \omega$ of flat space. The 
next step is to work out the Newtonian approximation for the gravitational 
self-interaction of this 0-point energy. Comparison with the known perturbative
result indicates how to include relativistic effects. Then the analysis is 
generalized for an arbitrary homogeneous and isotropic background. 

Let us recall some facts about the classical background:
\begin{equation}
ds^2_{\rm class} = -dt^2 + e^{2 H t} d{\vec x} \cdot d{\vec x} \; .
\end{equation}
Because it is not physically sensible to assume that coherent inflation 
commences over a region of more than about one Hubble volume, we work on the 
manifold $T^3 \times \Re$, where each of the coordinate radii is $H^{-1}$. The 
3-volume of this manifold is finite but expands exponentially:
\begin{equation}
V(t) = H^{-3} e^{3 H t} \; .
\end{equation}
By setting $ds^2_{\rm class} = 0$ we find the world line of a light ray which 
passes through ${\vec x} = 0$, directed along the unit vector ${\widehat r}$ at
$t = t_0$:
\begin{equation}
{\vec r}(t) = {{\widehat r} \over H} \left( e^{-H t_0} - e^{- H t}\right) \; .
\end{equation}
Multiplying by $\exp(H t)$ and taking the norm gives the physical distance from
the origin along the surface of simultaneity:
\begin{equation}
e^{H t} \Vert {\vec r}(t) \Vert = {1 \over H} \left(1 - e^{-H (t_0 - t)}\right)
\; .
\end{equation}
From this we infer the existence of a causal horizon of physical distance
$H^{-1}$, beyond which even a signal traveling at the speed of light can 
never reach. We can also compute the invariant 4-volume of the past light cone
from the point $(t,{\vec x}) = (t_0,0)$ to the surface of simultaneity at 
$t=0$:
\begin{eqnarray}
P_{\rm class}(t_0) & = & \int_0^{t_0} dt \; e^{3 H t} \times \frac43 \pi \Vert 
{\vec r}(t) \Vert^3 \; , \\
& = & \frac43 \pi H^{-4} \left(H t_0 - \frac{11}6 + 3 e^{-H t_0} - \frac32 
e^{-2 H t_0} + \frac13 e^{-3 H t_0}\right) \; . \label{eq:vplc}
\end{eqnarray}
Finally, note that the coordinate transformation $\eta = -H^{-1} \exp(-H t)$
makes the classical background conformal to flat space:
\begin{equation}
ds^2_{\rm class} = \Omega^2 \left(-d\eta^2 + d{\vec x} \cdot d{\vec x}\right)
\; , \label{eq:conformal}
\end{equation}
where the conformal factor is $\Omega \equiv -1/(H \eta)$. Note that the 
surface of simultaneity at $t=0$ corresponds to $\eta = -1/H$, and that the 
infinite future corresponds to $\eta \rightarrow 0^-$.

The next step is the kinematics of free gravitons. Graviton modes are described
by a polarization and by a co-moving wave number of the form ${\vec k} = 2 \pi 
H {\vec n}$, where ${\vec n}$ is a 3-tuple of integers. The integral 
approximation to a mode sum is:
\begin{equation}
2 \sum_{\vec n} \; f\left(2\pi H {\vec n}\right) \approx 2 \int d^3n \; 
f\left(2 \pi H {\vec n}\right) = 2 \int {d^3k \over (2 \pi H)^3} \; f({\vec k})
\; ,
\end{equation}
where the infrared cutoff is at $k \equiv \Vert {\vec k} \Vert = H$. Since
physical distances expand by $\exp(Ht)$, physical wave numbers redshift by 
$\exp(-Ht)$. We are most interested in infrared modes, defined as those whose
physical wave lengths have expanded beyond the causal horizon:
\begin{equation}
{\rm Infrared} \qquad \Longleftrightarrow \qquad H \ltwid k \ltwid H e^{Ht}
\; . \label{eq:defIR}
\end{equation}
We shall refer to the higher modes as ``ultraviolet.''

Now consider the dynamics of free gravitons. For any homogeneous and isotropic
geometry, these are the same as those of a massless, minimally coupled scalar
\cite{grish1}. Suppose we call such a field $\psi(\eta,{\vec x})$. In the
classical background its Lagrange density is:
\begin{equation}
{\cal L} = \frac12 \Omega^2 \left({\psi'}^2 - {\vec \nabla} \psi \cdot {\vec
\nabla} \psi\right) \; ,
\end{equation}
where a prime denotes differentiation with respect to the conformal time 
$\eta$. The mode coordinates are obtained by taking the spatial Fourier 
transform and multiplying by a factor of $H$:
\begin{equation}
q_{\vec k}(\eta) \equiv H \int d^3x \; e^{i {\vec k} \cdot {\vec x}} \psi(\eta,
{\vec x}) \; . \label{eq:modecoord}
\end{equation}
These variables allow us to recognize the Lagrangian as a sum of independent
harmonic oscillators:
\begin{equation}
L \equiv \int d^3 {\cal L} = \frac12 H \Omega^2 \sum_{\vec k} \left( 
q_{\vec k}^{\prime*} q_{\vec k}^{\prime} - k^2 q_{\vec k}^* \: 
q_{\vec k}\right) \; . \label{eq:mmslag}
\end{equation}
Since $q_{- {\vec k}} = q_{\vec k}^*$ we can treat this system as if there were
a single real mode for each wave number ${\vec k}$.
 
It is straightforward to express the mode coordinate and its conjugate momentum
in terms of creation and annihilation operators. The Heisenberg equation of 
motion is:
\begin{equation}
q^{\prime\prime}_{\vec k} - {2 \over \eta} q^{\prime}_{\vec k} + k^2 q_{\vec k}
= 0 \; .
\end{equation}
It follows that the negative frequency mode solution is:
\begin{equation}
u(\eta,k) = {\Omega^{-1} \over \sqrt{2 k}} \left(1 - {i \over k \eta}\right)
e^{-i k \eta} \; ,
\end{equation}
and we define its time derivative as:
\begin{equation}
u^{\prime}(\eta,k) \equiv -i k \Omega^{-2} v(\eta,k) = -ik {\Omega^{-1} \over 
\sqrt{2 k}} e^{- i k \eta} \; .
\end{equation}
The Wronskian formed from $u(\eta,k)$ and $v(\eta,k)$ is constant in 
consequence of the equation of motion, and with our normalization its value is:
\begin{equation}
u(\eta,k) v^*(\eta,k) + u^*(\eta,k) v(\eta,k) = {1 \over k} \; .
\end{equation} 
We can express $q_{\vec k}(\eta)$ as a linear combination of $u(\eta,k)$ and 
$u^*(\eta,k)$:
\begin{equation}
q_{\vec k}(\eta) = u(\eta,k) a_{\vec k} + u^*(\eta,k) a^{\dagger}_{\vec k} \; .
\end{equation}
The conjugate momentum is:
\begin{eqnarray}
p_{\vec k}(\eta) & = & H \Omega^2 q^{\prime}_{\vec k}(\eta) \; , \\
& = & -i k H v(\eta,k) a_{\vec k} + i k H v(\eta,k) a^{\dagger}_{\vec k} \; .
\end{eqnarray}
Requiring that $q_{\vec k}(\eta)$ commute canonically with $p_{\vec k}(\eta)$
and making use of the Wronskian determines the commutator of $a_{\vec k}$ and 
$a^{\dagger}_{\vec k}$ to be:
\begin{equation}
[a_{\vec k},a^{\dagger}_{\vec k}] = {1 \over H} \; .
\end{equation}

The ``Hamiltonian'' which generates the conformal time evolution of mode ${\vec
k}$ is:
\begin{equation}
H^{\eta}_{\vec k} = {p^2_{\vec k} \over 2 H \Omega^2} + \frac12 H k^2 \Omega^2 
q^2_{\vec k} \; .
\end{equation}
This is just a harmonic oscillator with frequency $k$ and mass $H \Omega^2$.
Because the mass is time dependent there are no stationary states but one can 
of course compute the expectation value of $H^{\eta}_{\vec k}$ in the presence 
of some state. In the far ultraviolet curvature is obviously a small effect, so
we may assume the flat space vacuum:
\begin{equation}
a_{\vec k} \vert 0 \rangle = 0 \; . \label{eq:vacuum}
\end{equation}
Note that Heisenberg states do not evolve, and that the operator $a_{\vec k}$ 
was constructed to be time independent. Hence condition (\ref{eq:vacuum}) 
persists, {\it even after the originally ultraviolet mode has red shifted to 
the infrared.} 

The expectation value of $H_{\eta}$ is simple to take in the presence of this 
state:
\begin{eqnarray}
\langle 0 \vert H^{\eta}_{\vec k}(\eta) \vert 0 \rangle & = & {1 \over 2 H 
\Omega^2} \times H^2 k^2 v(\eta,k) v^*(\eta,k) \times {1 \over H} \nonumber \\
& & \hskip 2cm + \frac12 H k^2 \Omega^2 \times u(\eta,k) u^*(\eta,k) \times 
{1 \over H} \; , \\
& = & \frac14 k + \frac14 k \left(1 + {1 \over k^2 \eta^2}\right) \; , \\
& = & \frac12 k + {1 \over 4 k \eta^2} \; .
\end{eqnarray}
Now exploit the relation between co-moving time and conformal time to relate
the co-moving Hamiltonian to the conformal one:
\begin{equation}
H^{t} = i {\partial \over \partial t} = i \Omega^{-1} {\partial \over \partial 
\eta} = \Omega^{-1} H^{\eta} \;
\end{equation}
It follows that the physical energy in mode ${\vec k}$ at co-moving time $t$
is:
\begin{equation}
E_{\vec k} = \frac12 k e^{-H t} + {H^2 \over 4 k} e^{H t}
\end{equation}
The first term is just the familiar 0-point energy, appropriately red shifted. 
One way to understand the second term is that virtual gravitons whose physical 
wave lengths exceed the Hubble radius cannot recombine; they are pulled apart 
by the expansion of spacetime. The energy of any one such graviton redshifts, 
but there are so many produced that the total energy contributed by each 
infrared mode actually increases.

Since there are an infinite number of ultraviolet modes, the total 0-point 
energy diverges. This is not consistent with the assumption that the background
is initially undergoing inflation with Hubble constant $H$. To make the 
assumption consistent we must subtract the original 0-point energy by normal 
ordering. However, this has only a minuscule effect on the superadiabatically 
amplified 0-point energy of the infrared modes. We can use simple Newtonian 
ideas to obtain a crude estimate of the energy density induced by their 
self-gravitation.

To obtain the energy density of mode ${\vec k}$ we divide by the 3-volume:
\begin{equation}
\rho_{\vec k} = \left(E_{\vec k} - \frac12 k e^{-H t} \right) \div V(t) =
{H^5 \over 4 k} e^{-2Ht} \; . \label{eq:rhok}
\end{equation}
Although this red shifts to zero, it does so more slowly than pure radiation. 
One consequence is that the associated Newtonian potential remains constant:
\begin{equation}
- e^{-2Ht} k^2 \varphi_{\vec k} = 4 \pi G \rho_{\vec k} \qquad \Longrightarrow
\qquad \varphi_{\vec k} = - {\pi G H^5 \over k^3} \; .
\end{equation}
The total Newtonian potential from all infrared modes is accordingly:
\begin{equation}
\varphi_{\rm IR} = {1 \over \pi^2 H^3} \int_{H}^{H \exp[Ht]} dk k^2 
\varphi_{\vec k} = -{G H^2 \over \pi} Ht \; .  
\end{equation}
This combines with the total 0-point energy density of infrared modes:
\begin{equation}
\rho_{\rm IR} = {1 \over \pi^2 H^3} \int_{H}^{H \exp[H t]} dk k^2 \rho_{\vec k}
= {H^4 \over 8 \pi^2} \; ,
\end{equation}
to produce an {\it increasing} gravitational interaction energy density:
\begin{equation}
\rho_{\rm Newt}(t) = \varphi_{\rm IR} \cdot \rho_{\rm IR} = - {G H^6 \over 8 
\pi^3} H t \; . \label{eq:Newt}
\end{equation}
This is down from $\rho_{\rm IR}$ by a factor of the small number 
$G \Lambda/(3 \pi) \ltwid 10^{-12}$, but its time dependence eventually makes 
it the more important effect.

The Newtonian estimate we have just obtained compares fairly well with the 
exact result of the lowest non-trivial order in perturbation theory \cite{tw1}:
\begin{equation}
\rho(t) = - {G H^6 \over 8 \pi^3} \left\{(H t)^2 + O(H t) \right\} + O(G^2)
\; . \label{eq:truerho}
\end{equation}
The extra factor of $Ht$ derives from the inclusion of four relativistic 
effects which were omitted in the Newtonian approximation:
\begin{enumerate}
\item There is a 0-point pressure in addition to the 0-point energy density.
\item The Newtonian potential is not the only gravitational field.
\item The various gravitational potentials can carry momentum.
\item The gravitational interaction is not linear.
\end{enumerate}
Although one must really do the quantum field theoretic calculation to get the
right answer, it is simple enough to indicate how each of these effects 
modifies the Newtonian estimate.

If we assume that the 0-point stress-energy of each mode is separately 
conserved then the 0-point pressure works out to be:
\begin{equation}
p_{\vec k} = -\frac13 \rho_{\vec k} = - {H^5 \over 12 k} e^{-2 H t} \; .
\end{equation}
Like the 0-point energy density, the total 0-point pressure is only a constant
--- and very small --- fraction of the pressure in the cosmological constant.
However, unlike the 0-point energy, the 0-point pressure serves as the source 
for a gravitational potential whose homogeneous equation of motion is that of a 
massless, minimally coupled scalar \cite{tw3}. It is straightforward to compute
the retarded Green's function for this potential:
\begin{eqnarray}
G\left(\eta,{\vec x};\eta^{\prime},{\vec x}^{\prime}\right) & = & -2 \theta(
\eta - \eta^{\prime}) \int {d^3k \over (2\pi)^3} \; e^{{\vec k} \cdot ({\vec x}
- {\vec x}^{\prime})} {\rm Im}\left[u^*(\eta,k) \: u(\eta^{\prime},k)\right] 
\; ,\\
& = & -{\theta(\Delta \eta) \over 4 \pi} \left\{ {\Omega^{-1} {\Omega^{
\prime}}^{-1} \over \Delta x} \delta\left({\Delta \eta} - {\Delta x}\right) 
+ H^2 \theta\left({\Delta \eta} - {\Delta x}\right)\right\} , \;\;\;\;\;\;
\end{eqnarray}
where ${\Delta \eta} \equiv \eta - \eta^{\prime}$ and ${\Delta x} \equiv \Vert
{\vec x} - {\vec x}^{\prime}\Vert$. The first term is just like its flat space 
cognate: the only contribution comes from sources on the actual light cone, so
there is no growth for a constant source. However, the second term superposes 
over the entire past light cone, whose invariant volume (\ref{eq:vplc}) grows 
like $H t$ in the classical background. 

The third relativistic effect means that we should not view the precipitating
event as the creation of two gravitons with opposite 3-momenta, whose stress 
energy then induces a gravitational potential containing zero 3-momentum. What 
really happens is the creation of a graviton with 3-momentum ${\vec k}_1$ and 
another with 3-momentum ${\vec k}_2$, which together induce a potential with 
3-momentum $-({\vec k}_1 + {\vec k}_2)$. Although this still leaves two mode 
sums, they are not cleanly split between a 0-point stress energy and a 
potential term, as was the case for our Newtonian estimate. 

The final relativistic effect means that we must include a bewildering variety
of interactions where the potential scatters off one of the gravitons, or
where it interacts with itself. This is one of the things that makes the
quantum field theoretic calculation so difficult. We can nonetheless say that
the effect is still due to the gravitational interaction between virtual 
gravitons which are pulled apart by inflation. Superadiabatic amplification can
still be used to estimate the rate at which these gravitons are created and the
stress energy which they carry. And it is generally the case that one factor of
$Ht$ derives from one of the two mode sums while the other factor of $Ht$ comes
from an integration over interaction times.

Correcting the Newtonian estimate for the induced energy density is not quite
the end. Just as a pressure is associated with $\rho_k$, so there is a pressure
associated with $\rho(t)$. We can find it by using energy conservation, which
reads as follows in the classical background (\ref{eq:classical}):
\begin{equation}
\dot{\rho} = - 3H (\rho + p) \; .
\end{equation}
Combined with (\ref{eq:truerho}) this implies that most rapidly growing part of
the induced pressure is exactly opposite that of the energy density:
\begin{equation}
p(t) = + {G H^6 \over 8 \pi^3} \left\{(H t)^2 + O(Ht)\right\} + O(G^2) \; .
\label{eq:truep}
\end{equation}
It follows that the interaction between infrared gravitons acts to screen
inflation by an amount that becomes non-perturbatively large at late times.

At this point it is useful to recall some standard facts about inflation
\cite{Linde,kt} in order to form a proper impression both of the effect's
magnitude and of the time scale over which it acts. What is usually termed, the
``scale of inflation,'' is the mass $M$ defined so that $M^4$ equals the energy
density of the cosmological constant, $\Lambda/(8\pi G)$. Since the Planck mass
is $M_p = G^{-1/2}$ we have:
\begin{equation}
G \Lambda = 8 \pi \left({M \over M_P}\right)^4 \; .
\end{equation}
It is traditional to assume that $M$ is GUT scale, which makes $G \Lambda
\sim 10^{-11}$. One sometimes encounters models with scales as low as that
of electroweak symmetry breaking. (Past that point there is not enough CP 
violation to explain the observed baryon asymmetry.) Inflation on the 
electroweak scale would give $G \Lambda \sim 10^{-67}$. These numbers mean 
that the gravitational interaction energy density is a very small faction 
of $M^4$ unless $Ht$ is enormous.

Although there are higher order corrections to the induced energy density 
(\ref{eq:truerho}) and pressure (\ref{eq:truep}), it turns out that the lowest 
order effect becomes non-perturbatively strong when these higher terms are 
still insignificant \cite{tw4}. The way this works is that the induced stress 
tensor serves as a source for corrections to the classical background 
(\ref{eq:classical}). The pressure again engenders an extra factor of $H t$
from the invariant volume of the past light cone \cite{tw3}, and this causes 
the lowest order effect to throttle inflation before higher orders can become 
significant. When the background is expressed in co-moving coordinates 
(\ref{eq:element}), the function $b(t)$ has the form:
\begin{equation}
b(t) = Ht [1 + D(t)] + {1 \over 2} \ln[1 + A(t)] \;, \label{eq:b(t)}
\end{equation}
where we recall from (\ref{eq:D}-\ref{eq:A}) the perturbative expansions of 
$D(t)$ and $A(t)$ for small $\epsilon \equiv G \Lambda/(3 \pi)$ and large $H t$
\cite{tw4}:
\begin{eqnarray}
D(t) & = & +{19 \over 2} (\epsilon H t)^2 + O\left((\epsilon H t)^3\right) 
\; , \\
A(t) & = & -{172 \over 9} \epsilon^2 (H t)^3 + O\left(\epsilon^3 (H t)^4\right)
\; ,
\end{eqnarray}
Perturbation theory breaks down when $A(t)$ approaches $-1$, at which time
only the first term in the expansion of $A(t)$ is relevant and no term in 
$D(t)$ is significant \cite{tw4}. 

The function $A(t)$ is of great importance because we saw, at the close of 
Section 2, that it gives the scalar during the perturbative regime:
\begin{equation}
{\widetilde \phi}(t) = - {1 \over \sqrt{8 \pi G}} \ln\left[1 + A(t)\right] \; .
\end{equation}
Summarizing and abstracting the preceding analysis, we may conclude that 
$A(t)$ derives, during the perturbative regime, from acting the retarded scalar
Green's function, ${\widetilde {\Box}}^{-1}$, on a source that grows like 
$(H t)^2$:
\begin{equation}
A =  -8\pi G \int_0^t d\tp e^{-3H\tp} \int_0^{\tp} d\tpp e^{3 H \tpp} {\rm 
Source}(\tpp) = 8 \pi G {1 \over {\widetilde {\Box}}} \left({\rm Source}\right)
\; , \label{eq:Adepsource}
\end{equation}
\begin{equation}
{\rm Source} = 172 {G H^6 \over 8 \pi^3} (H t)^2 + \dots \; . 
\end{equation}
This source is the stress energy induced by the gravitational interaction 
between gravitons produced by superadiabatic amplification. It consists of a
variety of stress energy-potential and potential-potential terms, each of which
contains two mode sums over the 3-momenta of the gravitons. One factor of $H t$
is attributable to one of these mode sums, the other comes from the 
superposition over interaction times in a potential. Part of the task of 
generalization is straightforward since the potentials are obtained from the 
retarded Green's function, which can be defined for any homogeneous and 
isotropic background:
\begin{equation}
{1 \over {\widehat {\Box}}} f \equiv - \int_0^t d\tp e^{-3 b(\tp)} \int_0^{\tp}
d\tpp e^{3 b(\tpp)} f(\tpp) \; .
\end{equation}
What remains is to understand superadiabatic amplification on a general 
homogeneous and isotropic background.

We begin by comparing the co-moving element with the conformal one:
\begin{equation}
d{\widehat s}^2 = -dt^2 + e^{2 b(t)} d{\vec x} \cdot d{\vec x} = \Omega^2 
\left(-d\eta^2 + d{\vec x} \cdot d{\vec x}\right) \; ,
\end{equation}
to infer general relations for the conformal factor and the coordinate 
transformation:
\begin{equation}
\Omega(\eta) = \exp[b(t)] \hskip 2cm dt = \Omega d\eta \; . \label{eq:transf}
\end{equation}
Let us denote by $\eta = \eta_i$ the image of $t=0$. Although we cannot be
explicit for a general $b(t)$, any superluminally expanding spacetime will have
$\eta_i < 0$. We can also assume that the approach to the infinite future is
$\eta \rightarrow 0^-$, as before. The manifold is still $T^3 \times \Re$, with 
each of the coordinate radii equal to $H^{-1}$. We therefore conclude that the 
3-volume and the graviton mode sum are:
\begin{eqnarray}
V(t) & = & H^{-3} e^{3 b(t)} \; ,\\
2 \sum_{\vec n} f\left(2 \pi H \Vert {\vec n}\Vert \right) & \approx & {1 \over 
\pi^2 H^3} \int dk k^2 f(k) \; .
\end{eqnarray}
It turns out that there is always a factor of $V^{-1}$ associated with each
mode sum so that the factors of $H^3$ cancel and any physical dependence upon 
the range of co-moving coordinates must come from the limits of integration.

Dynamical graviton modes still obey the equation of motion of a massless, 
minimally coupled scalar \cite{grish1}. The mode coordinates for such a scalar 
are still obtained by spatial Fourier transforming according to 
(\ref{eq:modecoord}), and the formula for the Lagrangian is unchanged from 
(\ref{eq:mmslag}), provided the general conformal factor is understood. What 
changes is the Heisenberg equations of motion:
\begin{equation}
q^{\prime\prime}_{\vec k} + 2 {\Omega^{\prime} \over \Omega} q^{\prime}_{\vec 
k} + k^2 q_{\vec k} = 0 \; .
\end{equation}
Redefining the field variable as $Q(\eta) \equiv \Omega q_{\vec k}(\eta)$ gives
the following suggestive equation:
\begin{equation}
Q^{\prime\prime} + \left(k^2 - {\Omega^{\prime\prime} \over \Omega}\right) Q 
= 0 \; . \label{eq:suggestive}
\end{equation}
There are two regimes in which good approximate solutions can be found: the
far ultraviolet, where $k^2 \gg \Omega^{\prime\prime}/\Omega$, and the far 
infrared where $k^2 \ll \Omega^{\prime\prime}/\Omega$. Before going on to study
these regimes we should note the important fact that they can be given an
invariant specification. In view of the relation (\ref{eq:transf}) between 
conformal and co-moving coordinates we can write:
\begin{equation}
{\Omega^{\prime\prime} \over \Omega} = {d \over dt} e^{b} {d \over dt} e^{b} =
e^{2 b} \left(\ddot{b} + 2 \dot{b}^2\right) = \frac16 e^{2b} {\widehat R} \; .
\end{equation}
The two regimes are therefore characterized by how the physical (i.e., red 
shifted) momentum compares with the square root of the Ricci scalar. In view of
our definition (\ref{eq:defIR}) for the classical background, it is reasonable 
to make the following general definition for ``infrared'' gravitons:
\begin{equation}
{\rm Infrared} \qquad \Longleftrightarrow \qquad e^{-2 b(t)} H^2 \ltwid e^{-2 
b(t)} k^2 \ltwid \frac1{12} {\widehat R}(t) \; .
\end{equation}
As before, we refer to the higher modes as ``ultraviolet.''

In the ultraviolet regime the Ricci scalar term is a perturbation:
\begin{equation}
Q^{\prime\prime} + k^2 Q = {\Omega^{\prime\prime} \over \Omega} Q \; .
\end{equation}
We can obtain the mode functions by iterating those of flat space:
\begin{equation}
Q_{\rm uv}(\eta,k) = {1 \over \sqrt{2 k}} e^{-i k \eta} + \int_{-\infty}^{\eta} 
d{\overline \eta} \: {1 \over k} \sin\left[k(\eta - {\overline \eta})\right] \:
{\Omega^{\prime\prime}({\overline \eta}) \over \Omega({\overline \eta})} \:
Q_{\rm uv}({\overline \eta},k) \; . \label{eq:iteration}
\end{equation}
This obviously results in a series in powers of $1/k$, the successive terms of
which are less and less significant for large, negative $\eta$. 

The mode functions associated with $Q_{\rm uv}(\eta,k)$ are:
\begin{eqnarray}
u(\eta,k) & = & \Omega^{-1}(\eta) Q_{\rm uv}(\eta,k) \; ,\\
v(\eta,k) & = & {i \over k} \left[\Omega(\eta) Q^{\prime}_{\rm uv}(\eta,k) -
\Omega^{\prime}(\eta) Q_{\rm uv}(\eta,k)\right] \; .
\end{eqnarray}
The associated Wronskian:
\begin{eqnarray}
\lefteqn{u(\eta,k) v^*(\eta,k) + u^*(\eta,k) v(\eta,k) =} \nonumber \\
& & {i \over k} \left[Q^*_{\rm uv}(\eta,k) Q_{\rm uv}^{\prime}(\eta,k) - 
Q_{\rm uv}(\eta,k) Q_{\rm uv}^{\prime *}(\eta,k)\right]\; .
\end{eqnarray}
is constant in consequence of the equation of motion (\ref{eq:suggestive}).
Since $Q_{\rm uv}(\eta,k)$ approaches the flat space mode functions at early
times, we see that the constant is $1/k$, the same as before. The subsequent
operator expansions and commutation relations:
\begin{eqnarray}
q_{\vec k}(\eta) & = & u(\eta,k) a_{\vec k} + u^*(\eta,k) a^{\dagger}_{\vec k} 
\; , \\
p_{\vec k}(\eta) & = & -i k H v(\eta,k) a_{\vec k} + i k H v(\eta,k) 
a^{\dagger}_{\vec k} \; ,
\end{eqnarray}
\begin{equation}
\left[a_{\vec k},a^{\dagger}_{\vec k}\right] = {1 \over H} \; ,
\end{equation}
are the same as for the classical background. And the expectation value of the
Hamiltonian in the presence of the state annihilated by $a_{\vec k}$ is:
\begin{equation}
\langle 0 \vert H^{\eta}_{\vec k} \vert 0 \rangle = \frac12 \left[Q^{\prime}_{
\rm uv} - {\Omega^{\prime} \over \Omega} Q_{\rm uv}\right]^* \left[Q^{\prime}_{
\rm uv} - {\Omega^{\prime} \over \Omega} Q_{\rm uv}\right] + \frac12 k^2 Q^*_{
\rm uv} Q_{\rm uv} \; . \label{eq:generalE}
\end{equation}
Although this formula is correct for all times, obtaining the leading behavior
for small $\eta$ requires that we develop an infrared expansion for the mode
functions.

In the infrared regime the momentum term is a perturbation: 
\begin{equation}
Q^{\prime\prime} - {\Omega^{\prime\prime} \over \Omega} Q = -k^2 Q\; ,
\end{equation}
At zeroth order the two independent solutions are:
\begin{equation}
Q_{10}(\eta) = \Omega(\eta) \hskip 2cm Q_{20}(\eta) = - \Omega(\eta) \int_{
\eta}^0 {d{\overline \eta} \over \Omega^2({\overline \eta})} \; .
\end{equation}
For superluminal expansion the limit $\eta \rightarrow 0^-$ carries 
$Q_{10}(\eta)$ to infinity, while $Q_{20}(\eta)$ goes to zero as $\sim \eta 
\Omega^{-1}$. The limit $\eta \rightarrow - \infty$ has the opposite effect: 
$Q_{10}(\eta)$ approaches zero and $Q_{20}(\eta)$ goes to infinity as $\sim 
\eta \Omega^{-1}$. 

In view of the limiting forms for $Q_{10}(\eta)$ and $Q_{20}(\eta)$ it is the
advanced Green's function:
\begin{equation}
G_{\rm adv}(\eta,{\overline \eta}) = \theta({\overline \eta} - \eta) \left[
Q_{10}(\eta) Q_{20}({\overline \eta}) - Q_{20}(\eta) Q_{10}({\overline \eta})
\right] \; ,
\end{equation}
which gives a reasonable equation to iterate for the full solution based on
$Q_{20}$:
\begin{equation}
Q_2(\eta,k) = Q_{20}(\eta) - k^2 \int_{-\infty}^0 d{\overline \eta} \: G_{\rm
adv}(\eta,{\overline \eta}) Q_2({\overline \eta},k) \; .
\end{equation}
The resulting series is:
\begin{equation}
Q_{2} = \sum_{\ell=0}^{\infty} (-k^2)^{\ell} G^{\ell}_{\rm adv} \cdot Q_{20} 
\; , \label{eq:Q2exp}
\end{equation}
where the $\ell$-th ``power'' of the Green's function denotes the $\ell$-fold
integration:
\begin{equation}
G^{\ell}_{\rm adv} \cdot Q_{20} \equiv \int_{-\infty}^0 d\eta_1 \: G_{\rm adv}(
\eta,\eta_1) \cdots \int_{-\infty}^0 d\eta_{\ell} \: G_{\rm adv}(\eta_{\ell-1},
\eta_{\ell}) Q_{20}(\eta_{\ell}) \; 
\end{equation}
For asymptotically small $\eta$ this $\ell$-th power goes to zero like 
$\sim \eta^{2\ell} \Omega^{-1}$. The expansion (\ref{eq:Q2exp}) is therefore in
terms which are less and less significant at late times.

The square of $Q_{10}(\eta)$ is not integrable at $\eta=0$. To obtain the full 
solution based on $Q_{10}$ we must therefore begin iterating with a Green's 
function which is intermediate between advanced and retarded:
\begin{equation}
G_{\rm int}(\eta,{\overline \eta}) = \theta(\eta - {\overline \eta}) Q_{20}(
\eta) Q_{10}({\overline \eta}) + \theta({\overline \eta} - \eta) Q_{10}(\eta)
Q_{20}({\overline \eta}) \; .
\end{equation}
The result is a series whose terms have the asymptotic behavior $\sim 
\eta^{2\ell} \Omega$. For high enough $\ell$ the integral of such a term times 
$Q_{10}$ no longer converges at large $\eta$. At this point we must continue 
the iteration using the advanced Green's function. The full series is:
\begin{equation}
Q_1 = \sum_{\ell=0}^{N} (-k^2)^{\ell} G^{\ell}_{\rm int} \cdot Q_{10} +
\sum_{\ell=N+1}^{\infty} (-k^2)^{\ell} G^{\ell-N}_{\rm adv} \cdot G^N_{\rm int}
\cdot Q_{10} \; ,
\end{equation}
where $N$ is the order of perturbation theory at which we must change Green's
functions. As with the expansion for $Q_2$, the small $\eta$ behavior of each 
successive term is weaker than that of its predecessor by a factor of $\eta^2$.

Since $Q_1(\eta,k)$ and $Q_2(\eta,k)$ are independent solutions of the same
linear, second order differential equation as $Q_{\rm uv}(\eta,k)$, we can find
complex numbers $\alpha$ and $\beta$ to enforce the condition:
\begin{eqnarray}
Q_{\rm ir}(\eta,k) & \equiv & \alpha Q_1(\eta,k) + \beta Q_2(\eta,k) \; , \\
& = & Q_{\rm uv}(\eta,k) \; .
\end{eqnarray}
Although $Q_{\rm ir}(\eta,k) = Q_{\rm uv}(\eta,k)$ over the full range of 
$\eta$, we can estimate $\alpha$ by matching the zeroth order solutions at the 
boundary between ultraviolet and infrared. This gives the relation:
\begin{equation}
\alpha \sim {\Omega_1^{-1} \over \sqrt{2 k}} e^{-i k \eta_1} \; , 
\label{eq:estimate}
\end{equation}
where $\eta_1$ is the conformal time at which $2 k^2 = \Omega_1^{\prime\prime}/
\Omega_1$. 

Our estimate (\ref{eq:estimate}) can be checked whenever explicit solutions 
exist. For example, suppose the scale factor is a power law which obeys the 
same initial conditions ($b(0) = 0$ and $\dot{b}(0) = H$) as the classical
background:
\begin{equation}
e^{b(t)} = \left(1 + {H t \over p}\right)^p \; . \label{eq:powerlaw}
\end{equation}
(We assume $p > 1$ to make the expansion superluminal.) The conformal factor 
is:
\begin{equation}
\Omega = \left({\eta \over \eta_0}\right)^{\frac{p}{p-1}} \; ,
\end{equation}
where $\eta_0 = -p/(p-1)~H^{-1}$ and the properly normalized solution is:
\begin{equation}
Q(\eta,k) = \frac12 \sqrt{\pi \eta} e^{-i \frac{\pi}2 (\nu + \frac12)} 
H_{\nu}^{(2)}(k \eta) \quad , \quad \nu = \frac12 \left({3p-1 \over p-1}\right)
\; .
\end{equation}
The infrared solutions $Q_1(\eta,k)$ and $Q_2(\eta,k)$ give the two Bessel
functions from which the Hankel function is constructed:
\begin{eqnarray}
\alpha Q_1(\eta,k) & = & \frac{i}2 {\rm csc}(\nu \pi) e^{-\frac{i}2 (\nu +
\frac12)\pi} \sqrt{\pi \eta} J_{-\nu}(k \eta) \; , \\
\beta Q_2(\eta,k) & = & -\frac12 {\rm csc}(\nu \pi) e^{\frac{i}2 (\nu + 
\frac12)\pi} \sqrt{\pi \eta} J_{\nu}(k \eta) \; .
\end{eqnarray}
By making the first terms of the respective series expansions agree with 
$Q_{10}(\eta)$ and $Q_{20}(\eta)$ we infer the following expressions for 
$\alpha$ and $\beta$:
\begin{equation}
\alpha = {\Gamma(\nu) \over \sqrt{2 \pi k}} \left({-2i \over k \eta_0}\right)^{
\nu - \frac12} = {\Gamma(\nu) \over \sqrt{\pi}} \left(\frac1{32} - \frac18 
\nu^2\right)^{\frac14 - \frac12 \nu} {\Omega_1^{-1} \over \sqrt{2 k}} \; ,
\end{equation}
\begin{eqnarray}
\lefteqn{\beta = - i \nu \sqrt{{2 \pi \over k}} {{\rm csc}(\nu \pi) \over 
\Gamma(\nu+1)} \left({i k \eta_0 \over 2}\right)^{\nu - \frac12} } \nonumber \\
& & = -2 \nu i \sqrt{\pi} {{\rm csc}(\nu \pi) \over \Gamma(\nu+1)} 
\left(\frac1{32} - \frac18 \nu^2\right)^{\frac12 \nu - \frac14} {\Omega_1^{-1} 
\over \sqrt{2 k}} \; .
\end{eqnarray}
Except for powers very close to the superluminal bound ($p=1$) the index $\nu$
is of order one and we see that $\alpha$ agrees with estimate 
(\ref{eq:estimate}). Unless $\nu$ happens to be very close to an integer or 
half integer we also see that $\alpha$ and $\beta$ are comparable.

Our series expansions imply the following relations for late times:
\begin{eqnarray}
Q_{\rm ir}(\eta,k) & \rightarrow & \alpha \Omega(\eta) \; ,\\
Q^{\prime}_{\rm ir}(\eta,k) - {\Omega^{\prime}(\eta) \over \Omega(\eta)} Q_{\rm
ir}(\eta,k) & \rightarrow & \alpha \eta \Omega(\eta) \; .
\end{eqnarray}
Substitution in (\ref{eq:generalE}) gives the late time behavior of the 
(conformal) energy in mode ${\vec k}$:
\begin{equation}
\langle 0 \vert H^{\eta}_{\vec k}(\eta) \vert 0 \rangle \rightarrow \frac12
\alpha \alpha^* k^2 \Omega^2(\eta)  \sim {R_1 \over 12 k} \Omega^2(\eta) \; ,
\end{equation}
where $R_1$ is the value of the Ricci scalar at the time mode ${\vec k}$ 
crosses from the ultraviolet regime to the infrared. The form of $R_1(k)$ 
depends upon $b(t)$. For general power law inflation (\ref{eq:powerlaw}) the 
Ricci scalar is:
\begin{equation}
{\widehat R}(t) \equiv 6 \left(\ddot{b}(t) + 2 \dot{b}^2(t)\right) = 
\left({H \over 1 + Ht/p}\right)^2 \left(-{6 \over p} + 12\right) \; ,
\end{equation}
and we find:
\begin{equation}
R_1(k) = 12 \left({2p -1 \over 2p}\right)^{p-2 \over p-1} 
\left({H \over k}\right)^{2 \over p - 1} H^2 \; .
\end{equation}
Taking $p$ to infinity recovers the de Sitter result, $R_1(k) = 12 H^2$. We 
therefore expect that $R_1(k)$ is generally a slowly decreasing or constant 
function of $k$.

It is now straightforward to generalize what was done in classical background. 
The co-moving energy is down by a factor of $\Omega$ so it approaches:
\begin{equation}
E_{\vec k} \rightarrow \frac12 \alpha \alpha^* k^2 e^{b(t)} \sim {R_1 \over 48
k} e^{b(t)} \; .
\end{equation}
Dividing by the 3-volume gives the energy density:
\begin{equation}
\rho_{\vec k} \rightarrow \frac12 \alpha \alpha^* k^2 H^3 e^{-2b(t)} \sim
{H^3 R_1 \over 48 k} e^{-2 b(t)} \; . \label{eq:Rrho}
\end{equation}
If the stress energy in mode ${\vec k}$ is separately conserved it is 
straightforward to infer the pressure:
\begin{equation}
\dot{\rho}_{\vec k} = -3\dot{b} (\rho_{\vec k} + p_{\vec k}) \quad 
\Longrightarrow \quad p_{\vec k} = -\frac13 \rho_{\vec k} \; .
\end{equation}
We can even parallel our previous estimate for the Newtonian potential of mode 
${\vec k}$:
\begin{equation}
- e^{-2 b(t)} k^2 \varphi_{\vec k} = 4 \pi G \rho_{\vec k} \quad 
\Longrightarrow \quad \varphi_{\vec k} = - 2 \pi G \alpha \alpha^* H^3 \sim
- {\pi G R_1 H^3 \over 12 k^3} \; .
\end{equation}
Of course all these results reduce to those obtained for the classical 
background when we set $R_1 = 12 H^2$. The difference is that {\it we can now 
recognize the Ricci scalar as the source of the stress energy and the 
gravitational potentials whose various interactions lead to screening.}

It remains to see what becomes of the two mode sums. Although we have already
explained that the Newtonian estimate misses some essential features of the
fully relativistic result, it can still be used to get a rough idea. In the
Newtonian approximation the total infrared energy density and potential are the
following mode sums: \begin{eqnarray}
\rho_{IR} & = & {1 \over \pi^2 H^3} \int_H^{K(t)} dk k^2 \rho_{\vec k} \sim {1 
\over 48 \pi^2} e^{-2 b(t)} \int_H^{K(t)} dk k R_1(k) \; , \\
\varphi_{IR} & = & {1 \over \pi^2 H^3} \int_H^{K(t)} dk k^2 \varphi_{\vec k} 
\sim - {G \over 12 \pi} \int_H^{K(t)} {dk \over k} R_1(k) \; ,
\end{eqnarray}
where $K^2(t) \equiv {\widehat R}(t) \exp[2 b(t)]/12$. The point which
generalizes about this is that, for the dominant terms, one of the mode sums 
goes like $\exp[2b] k dk$, whereas the other has the form $dk/k$. 

The relation:
\begin{equation}
k^2 = {1 \over 12} R_1 e^{2 b_1} \; ,
\end{equation}
can be used to convert these mode sums into integrations over time. For 
example, the general power law inflation (\ref{eq:powerlaw}) gives:
\begin{equation}
k dk = \left({p-1 \over p}\right) {1\over 12} R_1 e^{2 b_1} \dot{b}_1 dt_1 \; .
\end{equation}
Except for very slow inflation ($p \sim 1$) the time dependence of the Ricci 
scalar is negligible compared with that of the scale factor. We can therefore 
write:
\begin{equation}
k dk \approx {1 \over 12} R_1 e^{2 b_1} \dot{b}_1 dt_1 \; . \label{eq:Rkdk}
\end{equation}
In the same approximation the total infrared energy density becomes:
\begin{equation}
\rho_{IR} \approx {1 \over 576 \pi^2} e^{-2 b(t)} \int_0^t d\tp \dot{b}(\tp)
e^{2 b(\tp)} {\widehat R}^2(\tp) \approx {{\widehat R}^2(t) \over 1152 \pi^2}
\; .
\end{equation}
Hence we conclude that the $\exp[2b] k dk$ mode sum engenders an extra factor 
of the Ricci scalar.

The $dk/k$ mode sum is more subtle. Of course we can convert to an integration
over time as before:
\begin{equation}
\varphi_{IR} \approx -{G \over 12 \pi} \int_0^t d\tp \dot{b}(\tp) 
{\widehat R}(\tp) \; .
\end{equation}
However, further progress requires that we neglect $\ddot{b}$ relative to 
$\dot{b}^2$. For the general power law (\ref{eq:powerlaw}) the ratio $\ddot{b}/
\dot{b}^2$ is $-1/p$, which suggests that the approximation is also valid for
rapid inflation. We can use this to introduce a second integration, and write
the result in terms of ${\widehat {\Box}}^{-1}$:
\begin{eqnarray}
\varphi_{IR} & \approx & - {G \over \pi} \int_0^t d\tp \dot{b}^3(\tp) \; , \\
& \approx & - {3 G \over \pi} \int_0^t d\tp e^{-3 b(\tp)} \int_0^{\tp} d\tpp
e^{3 b(\tpp)} \dot{b}^4(\tpp) \; \\
& \approx & {G \over 48 \pi} {1 \over {\widehat {\Box}}}\left({\widehat R}^2
\right) \; . \label{eq:Rdk/k}
\end{eqnarray}
Our conclusion is accordingly that the $dk/k$ mode sum goes to ${\widehat 
{\Box}}^{-1} R$.

It would be a mistake to take the Newtonian approximation too seriously. For
example, although the source term of equation (\ref{eq:Adepsource}) contains
two mode sums, they are not associated one with a 0-point stress energy and 
the other with a potential. Nor must the source have the Newtonian form of a
stress energy times a potential; one can also have products of derivatives of
potentials, and there are terms in which the two virtual gravitons interact 
with the potential at different times. What can reasonably be concluded is that
the source term of equation (\ref{eq:Adepsource}) involves two factors of 
$\Box^{-1}$ acting in some order on five Ricci scalars.

\section{Ans\"{a}tze for the scalar}

The purpose of this section is to discuss possibilities for the scalar 
$\phi[g]$ that are consistent with the results of the last three sections. We
begin by reviewing these results, then we tabulate the 33 candidate invariants 
which can be constructed from the retarded scalar Green's function $\Box^{-1}$ 
and the Ricci scalar $R$. The section closes with a discussion of the 
possibilities for involving other Green's functions, and for replacing some of 
the five Ricci scalars by perturbatively equivalent factors of $4 \Lambda$. 

Relation (\ref{eq:pertF}) of Section 2 suggests that we take the scalar to be
the logarithm of a dimensionless invariant:
\begin{equation}
\phi[g] = - {1 \over \sqrt{8 \pi G}} \ln(1 - F[g]) \; , \label{eq:phiF}
\end{equation}
In the classical background ({\ref{eq:classical}) this invariant must reduce 
to:\footnote{In fact ${\widetilde F}(t)$ is just minus the perturbative 
coefficient function $A(t)$ of expression (\ref{eq:A}).} 
\begin{equation}
{\widetilde F}(t) = {172 \over 9} \left({G \Lambda \over 3 \pi}\right)^2 
(H t)^3 + {\rm subdominant} \; , \label{eq:pertlimit}
\end{equation}
From Section 3 we learned that $F[g](x)$ must be a dimensionless invariant 
which depends causally upon the fields within the past lightcone of $x^{\mu}$. 
The requirement of a stable initial value problem implies that any factors of 
the Riemann tensor must be protected by at least two inverse derivatives, with
{\it derivatives} of the curvature protected by correspondingly more inverse 
derivatives. This is consistent with the fact that the important part of 
$F[g](x)$ must be non-local. However, correspondence with the known results for
$\Lambda = 0$ puts severe restrictions on how many inverse derivatives can
appear. To be precise, the induced stress tensor (\ref{eq:scalarT}) must either
vanish with $\Lambda$ or else it must not be more infrared singular than 
$\partial^4 \ln(\partial^2)$ when subjected to the derivative expansion for 
weak fields in a flat space background.

Section 4 used the physics of our mechanism to show that $F[g]$ has the form 
of the scalar retarded Green's function $\Box^{-1}$ acting on a source
constructed from two more factors of $\Box^{-1}$ and five factors of the Ricci
scalar $R$. This reduces the problem to combinatorics. There are 21 ways of 
placing the factors of $R$ when the two retarded Green's functions act in 
series, and there are 12 distinct placements when they act in parallel. Table~1
lists the 33 possibilities. If one requires that at least one factor of $R$ 
must reside at each of the three locations then only ten terms remain.

\begin{table}

\vbox{\tabskip=0pt \offinterlineskip
\def\tablerule{\noalign{\hrule}}
\halign to450pt {\strut#& \vrule#\tabskip=1em plus2em& \hfil#& \vrule#& 
\hfil#\hfil& \vrule#& \hfil#& \vrule#& \hfil#\hfil& \vrule#\tabskip=0pt\cr
\tablerule
\omit & height2pt & \omit && \omit && \omit && \omit &\cr
&&\omit\hidewidth \# &&\omit\hidewidth {\rm Candidate Source} \hidewidth&& 
\omit\hidewidth \#\hidewidth&& \omit\hidewidth {\rm Candidate Source}
\hidewidth&\cr
\omit & height2pt & \omit && \omit && \omit && \omit &\cr
\tablerule
\omit & height2pt & \omit && \omit && \omit && \omit &\cr
&& 1a && $2 \; R^5 \left(\Box^{-1} 1 (\Box^{-1} 1)\right)$
&& 1b && $R^5 (\Box^{-1} 1) \; (\Box^{-1} 1)$ &\cr
\omit & height2pt & \omit && \omit && \omit && \omit &\cr
\tablerule
\omit & height2pt & \omit && \omit && \omit && \omit &\cr
&& 2a && $2 \; R^4 \left(\Box^{-1} R (\Box^{-1} 1)\right)$
&& 2b && $R^4 (\Box^{-1} R) \; (\Box^{-1} 1)$ &\cr
\omit & height2pt & \omit && \omit && \omit && \omit &\cr
\tablerule
\omit & height2pt & \omit && \omit && \omit && \omit &\cr
&& 3a && $2 \; R^4 \left(\Box^{-1} 1 (\Box^{-1} R)\right)$
&& 3b && $R^3 (\Box^{-1} R^2) \; (\Box^{-1} 1)$ &\cr
\omit & height2pt & \omit && \omit && \omit && \omit &\cr
\tablerule
\omit & height2pt & \omit && \omit && \omit && \omit &\cr
&& 4a && $2 \; R^3 \left(\Box^{-1} R^2 (\Box^{-1} 1)\right)$
&& 4b && $R^3 (\Box^{-1} R) \; (\Box^{-1} R)$ &\cr
\omit & height2pt & \omit && \omit && \omit && \omit &\cr
\tablerule
\omit & height2pt & \omit && \omit && \omit && \omit &\cr
&& 5a && $2 \; R^3 \left(\Box^{-1} R (\Box^{-1} R)\right)$
&& 5b && $R^2 (\Box^{-1} R^3) \; (\Box^{-1} 1)$ &\cr
\omit & height2pt & \omit && \omit && \omit && \omit &\cr
\tablerule
\omit & height2pt & \omit && \omit && \omit && \omit &\cr
&& 6a && $2 \; R^3 \left(\Box^{-1} 1 (\Box^{-1} R^2)\right)$
&& 6b && $R^2 (\Box^{-1} R^2) \; (\Box^{-1} R)$ &\cr
\omit & height2pt & \omit && \omit && \omit && \omit &\cr
\tablerule
\omit & height2pt & \omit && \omit && \omit && \omit &\cr
&& 7a && $2 \; R^2 \left(\Box^{-1} R^3 (\Box^{-1} 1)\right)$
&& 7b && $R (\Box^{-1} R^4) \; (\Box^{-1} 1)$ &\cr
\omit & height2pt & \omit && \omit && \omit && \omit &\cr
\tablerule
\omit & height2pt & \omit && \omit && \omit && \omit &\cr
&& 8a && $2 \; R^2 \left(\Box^{-1} R^2 (\Box^{-1} R)\right)$
&& 8b && $R (\Box^{-1} R^3) \; (\Box^{-1} R)$ &\cr
\omit & height2pt & \omit && \omit && \omit && \omit &\cr
\tablerule
\omit & height2pt & \omit && \omit && \omit && \omit &\cr
&& 9a && $2 \; R^2 \left(\Box^{-1} R (\Box^{-1} R^2)\right)$
&& 9b && $R (\Box^{-1} R^2) \; (\Box^{-1} R^2)$ &\cr
\omit & height2pt & \omit && \omit && \omit && \omit &\cr
\tablerule
\omit & height2pt & \omit && \omit && \omit && \omit &\cr
&& 10a && $2 \; R^2 \left(\Box^{-1} 1 (\Box^{-1} R^3)\right)$
&& 10b && $(\Box^{-1} R^5) \; (\Box^{-1} 1)$ &\cr
\omit & height2pt & \omit && \omit && \omit && \omit &\cr
\tablerule
\omit & height2pt & \omit && \omit && \omit && \omit &\cr
&& 11a && $2 \; R \left(\Box^{-1} R^4 (\Box^{-1} 1)\right)$
&& 11b && $(\Box^{-1} R^4) \; (\Box^{-1} R)$ &\cr
\omit & height2pt & \omit && \omit && \omit && \omit &\cr
\tablerule
\omit & height2pt & \omit && \omit && \omit && \omit &\cr
&& 12a && $2 \; R \left(\Box^{-1} R^3 (\Box^{-1} R)\right)$
&& 12b && $(\Box^{-1} R^3) \; (\Box^{-1} R^2)$ &\cr
\omit & height2pt & \omit && \omit && \omit && \omit &\cr
\tablerule
\omit & height2pt & \omit && \omit && \omit && \omit &\cr
&& 13a && $2 \; R \left(\Box^{-1} R^2 (\Box^{-1} R^2)\right)$
&& \omit && \omit &\cr
\omit & height2pt & \omit && \omit && \omit && \omit &\cr
\tablerule
\omit & height2pt & \omit && \omit && \omit && \omit &\cr
&& 14a && $2 \; R \left(\Box^{-1} R (\Box^{-1} R^3)\right)$
&& \omit && \omit &\cr
\omit & height2pt & \omit && \omit && \omit && \omit &\cr
\tablerule
\omit & height2pt & \omit && \omit && \omit && \omit &\cr
&& 15a && $2 \; R \left(\Box^{-1} 1 (\Box^{-1} R^4)\right)$
&& \omit && \omit &\cr
\omit & height2pt & \omit && \omit && \omit && \omit &\cr
\tablerule
\omit & height2pt & \omit && \omit && \omit && \omit &\cr
&& 16a && $2 \; \left(\Box^{-1} R^5 (\Box^{-1} 1)\right)$
&& \omit && \omit &\cr
\omit & height2pt & \omit && \omit && \omit && \omit &\cr
\tablerule
\omit & height2pt & \omit && \omit && \omit && \omit &\cr
&& 17a && $2 \; \left(\Box^{-1} R^4 (\Box^{-1} R)\right)$
&& \omit && \omit &\cr
\omit & height2pt & \omit && \omit && \omit && \omit &\cr
\tablerule
\omit & height2pt & \omit && \omit && \omit && \omit &\cr
&& 18a && $2 \; \left(\Box^{-1} R^3 (\Box^{-1} R^2)\right)$
&& \omit && \omit &\cr
\omit & height2pt & \omit && \omit && \omit && \omit &\cr
\tablerule
\omit & height2pt & \omit && \omit && \omit && \omit &\cr
&& 19a && $2 \; \left(\Box^{-1} R^2 (\Box^{-1} R^3)\right)$
&& \omit && \omit &\cr
\omit & height2pt & \omit && \omit && \omit && \omit &\cr
\tablerule
\omit & height2pt & \omit && \omit && \omit && \omit &\cr
&& 20a && $2 \; \left(\Box^{-1} R (\Box^{-1} R^4)\right)$
&& \omit && \omit &\cr
\omit & height2pt & \omit && \omit && \omit && \omit &\cr
\tablerule
\omit & height2pt & \omit && \omit && \omit && \omit &\cr
&& 21a && $2 \; \left(\Box^{-1} 1 (\Box^{-1} R^5)\right)$
&& \omit && \omit &\cr
\omit & height2pt & \omit && \omit && \omit && \omit &\cr
\tablerule}}

\caption{Candidate sources for $F[g]$. To obtain the functional $F[g]$ act 
$- \frac{43}{48} \left(\frac{G}{12\pi}\right)^2 \Box^{-1}$ on the source.}

\end{table}

All of the terms in Table~1 are manifestly invariant, causal and non-local. 
Note also that various the factors of $R$ are protected by enough factors of 
$\Box^{-1}$ so as not to jeopardize the stability of the initial value problem.
There is no problem with the flat space limit since the source terms behave 
generically like $\partial^6$ in the weak field derivative expansion. This 
means that the scalar --- and hence the induced stress tensor --- goes like 
$\partial^4$, which is perfectly allowed. Though it would not have been 
legitimate to impose this as a requirement, it is reassuring to note that all
of the terms approach a constant for subluminal expansion.

It is important to note that the actual source need not consist of a single
term from Table~1; it could equally well be a linear combination. Though we 
will not attempt it here, an explicit {\it derivation} of the source seems
possible in two different ways. First, we might re-compute the leading 
two loop diagrams for a general homogeneous and isotropic background, using the
methods of Section 4 to isolate the dominant terms and to express them in 
invariant form. This would be difficult, but no more so than the original
computation. The second technique exploits the fact that the terms of Table~1
survive in the limit $\Lambda \rightarrow 0$. They must therefore appear in the
usual effective action,\footnote{This sounds paradoxical --- in view of the 
fact that the ``in'' vacuum does not evolve into the ``out'' vacuum --- but it 
isn't really. The distinction between the two vacua simply means that the 
imaginary part of the effective action is non-zero (in fact, infrared 
divergent) when evaluated at the classical background.} which can be invoked to
further constrain the allowed combinations. One might even be able to pick off 
the coefficients by subjecting the two loop effective action to an expansion in 
powers of curvature along the lines developed by Barvinsky and Vilkovisky 
\cite{Vilko}.

It remains to discuss two issues, the first of which is why the non-locality is
confined to inverses of the scalar d'Alembertian:
\begin{eqnarray}
\lefteqn{\Box = {1 \over \sqrt{-g}} \partial_{\mu} \left(\sqrt{-g} g^{\mu\nu} 
\partial_{\nu}\right) } \nonumber \\
& & \longrightarrow {\widehat {\Box}} = - e^{-3b} {d \over dt} \left(
e^{3 b} {d \over dt}\right) \longrightarrow {\widetilde {\Box}} = - e^{-3 H t}
{d \over dt} \left( e^{3 H t} {d \over dt}\right) \; .
\end{eqnarray}
Of course we {\it derived} its presence from the physics of the mechanism, but 
why do we not need to allow for the possibility of inverses of other 
differential operators? It has been pointed out that the kinematics of free
gravitons on a homogeneous and isotropic background is governed by ${\widehat
{\Box}}$ but there are also non-dynamical, constrained modes which possess a
different kinetic operator \cite{tw3}:
\begin{equation}
{\widehat D}_B \equiv -e^{-b} {d \over dt} \left(e^{-b} {d \over dt} e^{2 b}
\right) \; .
\end{equation}
We will additionally consider the kinetic operator of a massless, conformally 
coupled scalar:
\begin{equation}
{\widehat D}_{\rm conf} \equiv {\widehat {\Box}} - {1 \over 6} {\widehat R} = 
- e^{-2 b} {d \over dt} \left(e^b {d \over dt} e^b\right) \; ,
\end{equation}
although there is no dynamical reason to suspect its presence.

The reason we ignore ${\widehat D}_B^{-1}$ and ${\widehat D}_{\rm conf}^{-1}$ 
is that they act to produce constants during the perturbative regime, and they
are even less relevant for slower expansion. To see this, consider how each of
the three inverse differential operators acts on a function of time $f(t)$ for
an arbitrary homogeneous and isotropic background:
\begin{eqnarray}
{1 \over {\widehat {\Box}}} f & = & - \int_0^t d\tp \; e^{-3 b(\tp)} 
\int_0^{\tp} d\tpp \; e^{3 b(\tpp)} f(\tpp) \; , \\
{1 \over {\widehat D}_B} f & = & - e^{-2 b(t)} \int_0^t d\tp \; e^{b(\tp)} 
\int_0^{\tp} d\tpp \; e^{b(\tpp)} f(\tpp) \; , \\
{1 \over {\widehat D}_{\rm conf}} f & = & - e^{-b(t)} \int_0^t d\tp \; 
e^{-b(\tp)} \int_0^{\tp} d\tpp \; e^{2 b(\tpp)} f(\tpp) \; .
\end{eqnarray}
In the perturbative regime $b(t) = H t$ and ${\widetilde R}(t) = 12 H^2$. 
Therefore acting each of the three operators on the Ricci scalar gives:
\begin{eqnarray}
{1 \over {\widetilde {\Box}}} {\widetilde R} & = & - 4 H t + \frac43 - \frac43
e^{-3 H t} \; , \\
{1 \over {\widetilde D}_B} {\widetilde R} & = & - 6 \left(1 - 2 e^{-Ht} +
e^{-2Ht}\right) \; , \\
{1 \over {\widetilde D}_{\rm conf}} {\widetilde R} & = & - 6 \left(1 - 2 
e^{-Ht} + e^{-2Ht}\right) \; .
\end{eqnarray}
Except for the beginning of inflation, the operators ${\widehat D}_B$ and 
${\widehat D}_{\rm conf}$ are indistinguishable from constants during the 
perturbative regime! On the other hand, ${\widehat {\Box}^{-1}}$ grows like 
$H t$. When we recall that this factor must be at least $10^7$ before anything 
interesting happens, the other operators are negligible in comparison. 

The other operators perform even worse during the slower expansion that should 
follow the end of inflation. Suppose we call the transition time $t_z$. It is
useful to break the double integrations of the operators into periods which are
before and after $t_z$:
\begin{equation}
\int_0^t d\tp \int_0^{\tp} d\tpp = \int_0^{t_z} d\tp \int_0^{\tp} d\tpp +
\int_{t_z}^t d\tp \int_0^{t_z} d\tpp + \int_{t_z}^t d\tp \int_{t_z}^{\tp} d\tpp
\; . \label{eq:split}
\end{equation}
Since there are so many more e-foldings before $t_z$ than afterwards the 
largest contribution for each of the three operators comes from the first of 
the three integrals to the right of (\ref{eq:split}). Since this integral is
restricted to the perturbative regime we can estimate the post inflationary
behavior of the three operators:
\begin{eqnarray}
{1 \over {\widehat {\Box}}} {\widehat R} & \approx & - 4 H t_z \; , \\
{1 \over {\widehat D}_B} {\widehat R} & \approx & - 6 \: e^{-2 b(t) + 2 b(t_z)} 
\; , \\
{1 \over {\widehat D}_{\rm conf}} {\widehat R} & \approx & - 6 \: e^{-b(t) +
b(t_z)} \; .
\end{eqnarray}
Whereas ${\widehat {\Box}}^{-1}$ gives an enormous constant, the other 
operators are only of order unity at $t=t_z$, and they fall off thereafter. It
follows that these operators are irrelevant.

So much for other operators; the final issue is whether some of the five 
factors of $R$ should be replaced by $4 \Lambda$. Although these terms are 
distinct for a general homogeneous and isotropic background, it is not possible
to distinguish them during the perturbative regime:
\begin{equation}
R \longrightarrow {\widehat R} = 6 (\ddot{b} + 2 \dot{b}^2) \longrightarrow
{\widetilde R} = 12 H^2 \; . \label{eq:Ricci}
\end{equation}
For the factors of $R$ which occur furthest back in time it probably makes 
little difference whether or not they are replaced by $4 \Lambda$. The 
integrals are dominated by the inflationary period, during which the two terms 
agree. On the other hand, the factors of $R$ which appear latest might play an 
important role in the post inflationary regime where they are insignificant 
compared with $4 \Lambda$.

Of course one can consider the physics of screening on a general background, as
we did in Section 4. This favors $R$ over $4 \Lambda$. The preference is very 
strong for the two factors of $R$ from the 0-point energy of the virtual 
gravitons (\ref{eq:Rrho}) and for the factor from the $k dk$ mode sum
(\ref{eq:Rkdk}). It is somewhat weaker for the factor of $R$ from the $dk/k$ 
mode sum (\ref{eq:Rdk/k}), and it is very thin for the factor that was 
introduced to compensate the dimensions of the other $\Box^{-1}$ in the source. 

We can therefore find some justification to consider sources involving four or 
only three factors of the Ricci scalar. There are 15 ways of placing four $R$'s
when the two factors of $\Box^{-1}$ act sequentially, and there are 9 distinct 
placements for the parallel case. We shall not bother listing them all. Table~2
gives the 10 series and 6 parallel candidates that contain only three factors 
of $R$. It is worth remarking that the presence of explicit factors of
$\Lambda$ would preclude deriving the induced stress tensor from the usual
effective action, however, one could still derive the actual source by 
re-computing the dominant two loop contribution in a general homogeneous and
isotropic background.

\begin{table}

\vbox{\tabskip=0pt \offinterlineskip
\def\tablerule{\noalign{\hrule}}
\halign to450pt {\strut#& \vrule#\tabskip=1em plus2em& \hfil#& \vrule#& 
\hfil#\hfil& \vrule#& \hfil#& \vrule#& \hfil#\hfil& \vrule#\tabskip=0pt\cr
\tablerule
\omit & height2pt & \omit && \omit && \omit && \omit &\cr
&&\omit\hidewidth \# &&\omit\hidewidth {\rm Candidate Source} \hidewidth&& 
\omit\hidewidth \#\hidewidth&& \omit\hidewidth {\rm Candidate Source}
\hidewidth&\cr
\omit & height2pt & \omit && \omit && \omit && \omit &\cr
\tablerule
\omit & height2pt & \omit && \omit && \omit && \omit &\cr
&& 1c && $2 \; R^3 \left(\Box^{-1} 1 (\Box^{-1} 1)\right)$
&& 1d && $R^3 (\Box^{-1} 1) \; (\Box^{-1} 1)$ &\cr
\omit & height2pt & \omit && \omit && \omit && \omit &\cr
\tablerule
\omit & height2pt & \omit && \omit && \omit && \omit &\cr
&& 2c && $2 \; R^2 \left(\Box^{-1} R (\Box^{-1} 1)\right)$
&& 2d && $R^2 (\Box^{-1} R) \; (\Box^{-1} 1)$ &\cr
\omit & height2pt & \omit && \omit && \omit && \omit &\cr
\tablerule
\omit & height2pt & \omit && \omit && \omit && \omit &\cr
&& 3c && $2 \; R^2 \left(\Box^{-1} 1 (\Box^{-1} R)\right)$
&& 3d && $R (\Box^{-1} R^2) \; (\Box^{-1} 1)$ &\cr
\omit & height2pt & \omit && \omit && \omit && \omit &\cr
\tablerule
\omit & height2pt & \omit && \omit && \omit && \omit &\cr
&& 4c && $2 \; R \left(\Box^{-1} R^2 (\Box^{-1} 1)\right)$
&& 4d && $R (\Box^{-1} R) \; (\Box^{-1} R)$ &\cr
\omit & height2pt & \omit && \omit && \omit && \omit &\cr
\tablerule
\omit & height2pt & \omit && \omit && \omit && \omit &\cr
&& 5c && $2 \; R \left(\Box^{-1} R (\Box^{-1} R)\right)$
&& 5d && $(\Box^{-1} R^3) \; (\Box^{-1} 1)$ &\cr
\omit & height2pt & \omit && \omit && \omit && \omit &\cr
\tablerule
\omit & height2pt & \omit && \omit && \omit && \omit &\cr
&& 6c && $2 \; R \left(\Box^{-1} 1 (\Box^{-1} R^2)\right)$
&& 6d && $(\Box^{-1} R^2) \; (\Box^{-1} R)$ &\cr
\omit & height2pt & \omit && \omit && \omit && \omit &\cr
\tablerule
\omit & height2pt & \omit && \omit && \omit && \omit &\cr
&& 7c && $2 \; \left(\Box^{-1} R^3 (\Box^{-1} 1)\right)$
&& \omit && \omit &\cr
\omit & height2pt & \omit && \omit && \omit && \omit &\cr
\tablerule
\omit & height2pt & \omit && \omit && \omit && \omit &\cr
&& 8c && $2 \; \left(\Box^{-1} R^2 (\Box^{-1} R)\right)$
&& \omit && \omit &\cr
\omit & height2pt & \omit && \omit && \omit && \omit &\cr
\tablerule
\omit & height2pt & \omit && \omit && \omit && \omit &\cr
&& 9c && $2 \; \left(\Box^{-1} R (\Box^{-1} R^2)\right)$
&& \omit && \omit &\cr
\omit & height2pt & \omit && \omit && \omit && \omit &\cr
\tablerule
\omit & height2pt & \omit && \omit && \omit && \omit &\cr
&& 10c && $2 \; \left(\Box^{-1} 1 (\Box^{-1} R^3)\right)$
&& \omit && \omit &\cr
\omit & height2pt & \omit && \omit && \omit && \omit &\cr
\tablerule}}

\caption{Candidate sources for $F[g]$ with only three Ricci scalars. To obtain 
the functional $F[g]$ act $- \frac{43}{48} 
\left(\frac{G\Lambda}{3\pi}\right)^2 \Box^{-1}$ on the source.}

\end{table}

\section{Numerical evolution}

Despite the enormous restriction we have obtained on the possible forms for the
scalar $\phi[g]$, it is still not quite unique. The purpose of this section is 
to develop and to implement a scheme for numerically evolving whatever model is
defined when one finally settles upon a choice for $\phi[g]$. Our procedure 
will be to select one of the ans\"atze of the previous section and then evolve 
it far enough past the end of inflation to infer the asymptotic behavior at 
late times. We shall also obtain explicit results for the number of e-foldings 
which occur before the end of inflation, the rapidity of the transition, and 
for the equation of state throughout the evolution. It ought to be 
straightforward to generalize the numerical scheme to any of the viable 
ans\"atze discussed in Section 5.

Recall from Section 2 that the evolution equation is:
\begin{equation}
\ddot{b} = - 4 \pi G \left({d{\widehat \phi}[b] \over dt}\right)^2 \; ,
\end{equation}
subject to the initial condition $b(0) = 0$ and $\dot{b}(0) = H$. This gives
$b(t)$. From the field effective field equation (\ref{eq:rhotot}-\ref{eq:ptot})
we see that the total energy density and pressure, including the contribution
from the cosmological constant, are:
\begin{eqnarray}
\rho_{\rm tot} & = & {1 \over 8 \pi G} \: 3 \dot{b}^2 \; , \\
p_{\rm tot} & = & {1 \over 8 \pi G} \left(-2 \ddot{b} - 3 \dot{b}^2\right) \; .
\end{eqnarray}
This means that the instantaneous equation of state can be reconstructed as
follows:
\begin{equation}
{p_{\rm tot}(t) \over \rho_{\rm tot}(t)} = - {2 \ddot{b}(t) \over \dot{b}^2(t)}
- 1 \; .
\end{equation}

Recall as well that the scalar can be written in terms of a dimensionless 
invariant $F[g]$:
\begin{equation}
\phi[g] = - {1 \over \sqrt{8 \pi G}} \ln\left(1 - F[g]\right) \; .
\end{equation}
Substitution gives the following general evolution equation:
\begin{equation}
\ddot{b} = -\frac12 \left({d{\widehat F}/dt \over 1 - {\widehat F}[b]}
\right)^2 \; . \label{eq:Fform}
\end{equation}
In Section 5 we argued that $F[g]$ has the form of a retarded scalar Green's
function acting on a source composed of two more retarded scalar Green's 
functions which act in some order on three to five factors of the Ricci scalar.

Before specializing to a particular ans\"atz, we should comment on the 
qualitative behavior of all models. During inflation the time derivative of the
dynamical variable $b(t)$ is a large positive constant. Inflation is brought to
an end when the inherently negative second derivative becomes large enough to
force $\dot{b}(t)$ down to nearly zero. It is clear from (\ref{eq:Fform}) that 
this must occur when ${\widehat F}[b](t)$ approaches unity. All viable models 
show the following growth for ${\widetilde F}(t)$ during the perturbative 
regime:
\begin{equation}
{\widetilde F}(t) = \frac{172}3 \epsilon^2 (H t)^3 + {\rm subdominant} \; ,
\end{equation}
where we define the dimensionless parameter $\epsilon \equiv G\Lambda/(3\pi)$.
Our perturbative estimate for the number of inflationary e-foldings is 
accordingly \cite{tw4}:
\begin{equation}
N_{\rm pert} = \left({9 \over 172}\right)^{\frac13} \left({3 \pi \over G 
\Lambda} \right)^{-\frac23} = \left({81 \over 11008}\right)^{\frac13} 
\left({M_P \over M}\right)^{\frac83} \; , \label{eq:Npert}
\end{equation}
where $M_P$ is the Planck mass and $M$ is the scale of inflation. It is 
intriguing to plug in the numbers. For inflation on the GUT scale one has 
$N_{\rm pert} \sim 10^7$ e-foldings before screening becomes effective. For 
electroweak inflation we predict $N_{\rm pert} \sim 10^{45}$ e-foldings. 

For technical and historical reasons we chose to develop the scheme for source
4d of Table~2:
\begin{equation}
F[g] = -{43 \over 48} \epsilon^2 {1 \over \Box}\left(R \left({1 \over \Box} R
\right)^2 \right) \; . \label{eq:theansatz}
\end{equation}
The dimensionless parameter $\epsilon$ is only about $10^{-12}$, even for GUT 
scale inflation.\footnote{A minor point is that the parameter $\epsilon$ 
characterizes all models, not just those with three factors of the Ricci 
scalar. Even when the explicit factors of $\Lambda$ are replaced by $R$'s one 
still gets $\epsilon^2$ when the co-moving time is rescaled to the 
dimensionless variable $\tau \equiv H t$.} Specializing to a homogeneous and 
isotropic background we have:
\begin{equation}
{\widehat F}[b](t) = \frac{172}3 \epsilon^2 \int_0^t d\tp e^{-3 b(\tp)} 
\int_0^{\tp} d\tpp e^{3 b(\tpp)} \left( \frac32 \ddot{b}(\tpp) + 3 
\dot{b}^2(\tpp) \right) B^2[b](\tpp) \; ,
\end{equation}
where we define the functional $B[b](t)$ as follows:
\begin{equation}
B[b](t) \equiv \int_0^t d\tp e^{-3 b(\tp)} \int_0^{\tp} d\tpp e^{3 b(\tpp)}
\left( \frac32 \ddot{b}(\tpp) + 3 \dot{b}^2(\tpp) \right) \; .
\end{equation} 
The time derivative of ${\widehat F}[b](t)$ is simple to compute:
\begin{equation}
{d{\widehat F} \over dt} = \frac{172}3 \epsilon^2 \int_0^t d\tp e^{-3 b(t) + 3
b(\tp)} \left( \frac32 \ddot{b}(\tp) +3 \dot{b}^2(\tp) \right) B^2[b](\tp) \; .
\label{eq:dFhat/dt}
\end{equation}

Although the evolution equation we have obtained is both non-local and 
non-linear, it is simple to solve numerically by naive discretization. The 
independent variable $t$ is characterized by an integer $i$ and a dimensionless
step size $\Delta \tau$:
\begin{equation}
t \longrightarrow i H^{-1} {\Delta\tau} \; .
\end{equation}
Functions of $t$ become discrete in the usual way:
\begin{equation}
f(t) \longrightarrow f_i \; .
\end{equation}
Derivatives are discretized using the difference operator:
\begin{equation}
\dot{f}(t) \longrightarrow H \left({f_{i+1} - f_i \over \Delta\tau}\right) 
\equiv H {\Delta f_i \over \Delta\tau} \; .
\end{equation}
In order to preserve the fundamental theorem of integral calculus we must
discretize integrals by summing one step backwards:
\begin{equation}
\int_0^t d\tp f(\tp) \longrightarrow \sum_{j=0}^{i-1} H^{-1} {\Delta\tau} f_j
\: .
\end{equation}
It turns out that all the factors of $H^{-1} {\Delta\tau}$ cancel for our 
ans\"atz (this is its nice technical feature) and we obtain the following
discretized evolution equation:
\begin{equation}
{\Delta^2 b}_i = - \frac12 \left({{\Delta F}_i \over 1 - F_i}\right)^2 \; ,
\end{equation}
where the discrete versions of ${\widehat F}[b](t)$ and $B[b](t)$ are:
\begin{eqnarray}
F_i & \equiv & \frac{172}3 \epsilon^2 \sum_{j=0}^{i-1} e^{-3b_j} 
\sum_{k=0}^{j-1} e^{3 b_k} \left( \frac32 {\Delta^2 b}_k + 3 ({\Delta b}_k)^2 
\right) (B_k)^2 \; , \\
B_i & \equiv & \sum_{j=0}^{i-1} e^{-3b_j} \sum_{k=0}^{j-1} e^{3 b_k} 
\left( \frac32 {\Delta^2 b}_k + 3 ({\Delta b}_k)^2 \right) \; .
\end{eqnarray}
We have simplified the notation by dropping the hat on the discretized version 
of ${\widehat F}[b](t)$ because $F[g](t,{\vec x})$ and ${\widetilde F}(t)$ are
never discretized.

Although the discretization we have achieved is plausible, it would be tedious
to iterate on account of the need to compute summations at each step. This can
be avoided by simply keeping ${\Delta F}_i$ and ${\Delta B}_i$ as auxiliary 
variables. The resulting recursion scheme is:
\begin{eqnarray}
B_i & = & B_{i-1} + {\Delta B}_{i-1} \; , \\
{\Delta B}_i & = & e^{-3 {\Delta b}_{i-1}} \left({\Delta B}_{i-1} + \frac32 
{\Delta^2 b}_{i-1} + 3 ({\Delta b}_{i-1})^2 \right) \; , \\
F_i & = & F_{i-1} + {\Delta F}_{i-1} \; , \\
{\Delta F}_i & = & e^{-3 {\Delta b}_{i-1}} \left({\Delta F}_{i-1} + \frac{172}3
\epsilon^2 \left( \frac32 {\Delta^2 b}_{i-1} + 3 ({\Delta b}_{i-1})^2 \right) 
(B_{i-1})^2\right) \; , \;\;\;\;\;\; \\
b_i & = & b_{i-1} + {\Delta b}_{i-1} \; , \\
{\Delta b}_i & = & {\Delta b}_{i-1} + {\Delta^2 b}_{i-1} \; , \\
{\Delta^2 b}_i & = & -\frac12 \left({{\Delta F}_i \over 1 - F_i}\right)^2 \; .
\end{eqnarray}
Note that only differences of $b$ enter the scheme; the exponentials of $\pm 3 
b$ in the continuum theory are all absorbed into such terms. Note also that all 
the variables are initialized to zero except ${\Delta b}_0 = {\Delta \tau}$. 
It is only through this initial value that the scheme depends on the step size 
${\Delta \tau}$.

Although $\epsilon$ should be about $10^{-12}$ or less, the evolution is very 
slow for such small values. In this first study we accordingly ran the scheme 
for somewhat larger, although still sub-Planckian values: $\epsilon = 10^{-3}$,
$10^{-4}$, $10^{-5}$ and $10^{-6}$. The evolution was carried out with a step 
size of ${\Delta \tau} = 10^{-3}$ using the package Mathematica \cite{math}. 
The results for the effective Hubble constant $H_{\rm eff}(t) = \dot{b}(t)$ are
displayed in Figure~1. In each case the end of inflation is rapid and requires
about five e-foldings, in good agreement with the perturbative prediction 
\cite{tw4}.

\begin{figure}

\centerline{\psfig{figure=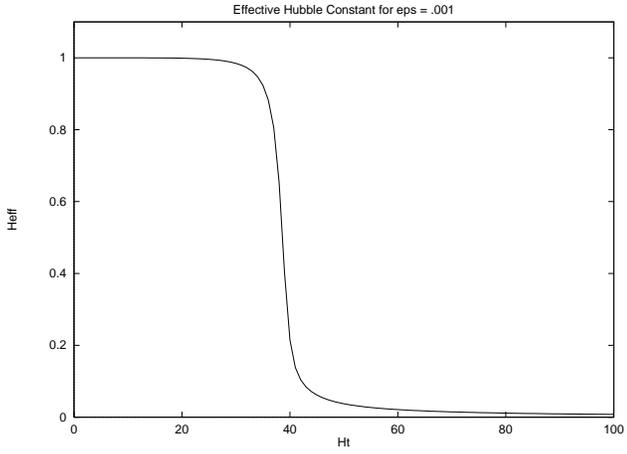,height=7cm,bbllx=0bp
,bblly=0bp,bburx=596bp,bbury=843bp,rheight=5.5cm,rwidth=10.5cm,angle=-90}
\hskip -1.5cm
\psfig{figure=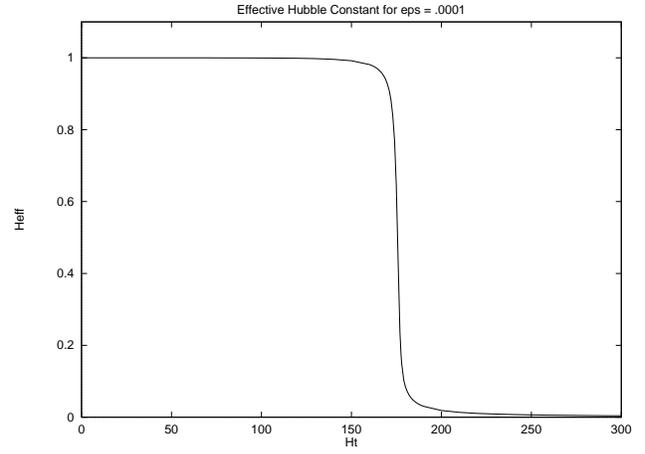,height=7cm,bbllx=0bp
,bblly=0bp,bburx=596bp,bbury=843bp,rheight=5.5cm,rwidth=10.5cm,angle=-90}}

\vskip 2cm

\centerline{\psfig{figure=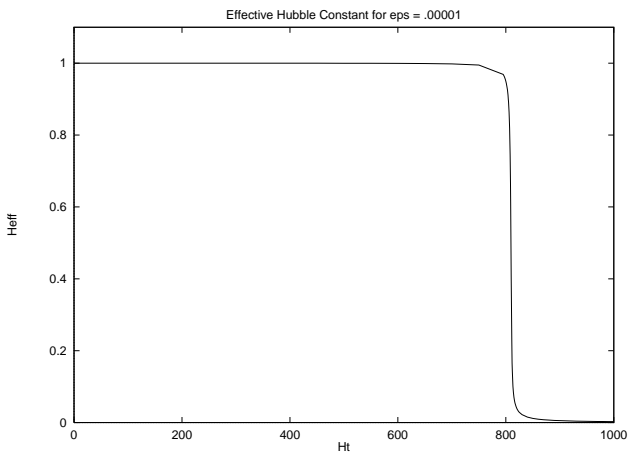,height=7cm,bbllx=0bp
,bblly=0bp,bburx=596bp,bbury=843bp,rheight=5.5cm,rwidth=10.5cm,angle=-90}
\hskip -1.5cm
\psfig{figure=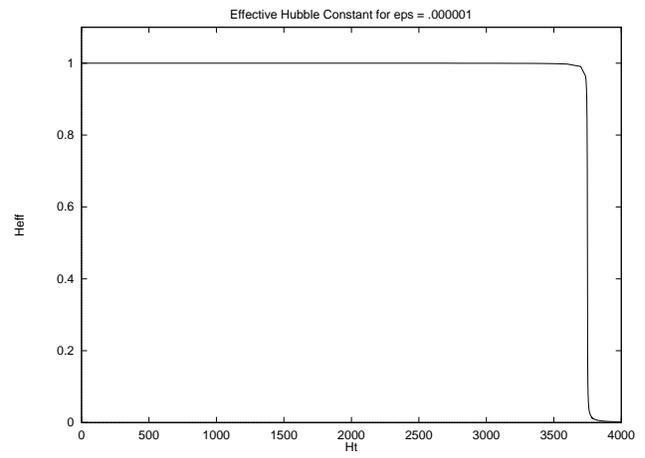,height=7cm,bbllx=0bp
,bblly=0bp,bburx=596bp,bbury=843bp,rheight=5.5cm,rwidth=10.5cm,angle=-90}}

\vskip 1cm

\caption{$H_{\rm eff}(t) \div H$ versus $Ht$ for $\epsilon = 10^{-3}$, 
$10^{-4}$, $10^{-5}$ and $10^{-6}$.}

\end{figure}

Figure~2 shows the instantaneous equation of state $p_{\rm tot}/\rho_{\rm tot}$
for the four values of $\epsilon$. The asymptotic equation of state is clearly
that of pure radiation: $p_{\rm tot} = \frac13 \rho_{\rm tot}$, corresponding 
to a scale factor which grows as the square root of the co-moving time. The 
reason for this is that the functional ${\widehat F}$ continues to approach one
as long as ${\widehat R} = 6 (\ddot{b} + 2 \dot{b}^2)$ is positive. As 
${\widehat F}$ approaches one, the second derivative of $b(t)$ become ever more
negative, which drives ${\widehat R}$ to zero. But this implies square root 
expansion:
\begin{equation}
{\widehat R}(t) = 0 \qquad \Longrightarrow \qquad \dot{b}(t) = {1 \over 2 (t -
t_z)} \; ,
\end{equation}
where the shift $t_z$ provides a reasonable definition for the time at which 
inflation ends.

\begin{figure}

\centerline{\psfig{figure=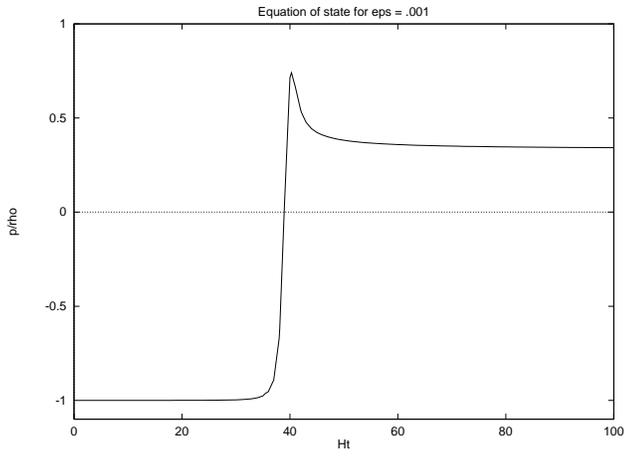,height=7cm,bbllx=0bp
,bblly=0bp,bburx=596bp,bbury=843bp,rheight=5.5cm,rwidth=10.5cm,angle=-90}
\hskip -1.5cm
\psfig{figure=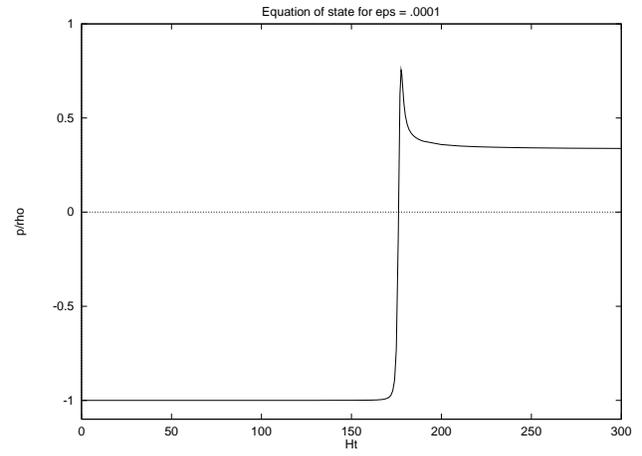,height=7cm,bbllx=0bp
,bblly=0bp,bburx=596bp,bbury=843bp,rheight=5.5cm,rwidth=10.5cm,angle=-90}}

\vskip 2cm

\centerline{\psfig{figure=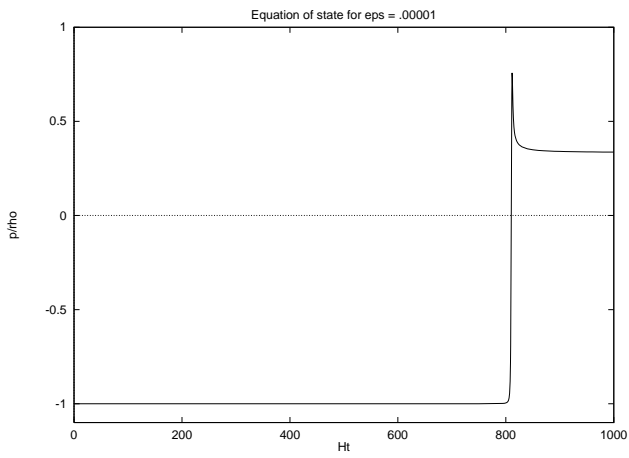,height=7cm,bbllx=0bp
,bblly=0bp,bburx=596bp,bbury=843bp,rheight=5.5cm,rwidth=10.5cm,angle=-90}
\hskip -1.5cm
\psfig{figure=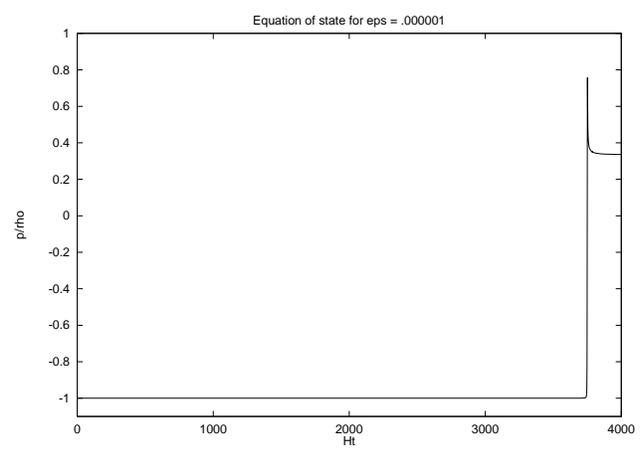,height=7cm,bbllx=0bp
,bblly=0bp,bburx=596bp,bbury=843bp,rheight=5.5cm,rwidth=10.5cm,angle=-90}}

\vskip 1cm

\caption{$p_{\rm tot}/\rho_{\rm tot}$ versus $Ht$ for $\epsilon = 10^{-3}$, 
$10^{-4}$, $10^{-5}$ and $10^{-6}$.}

\end{figure}

Of course the limiting form is approached asymptotically. The first corrections
in the series are:
\begin{eqnarray}
\dot{b}(t) & = & {1 \over 2 (t-t_z)} - {\alpha \ln\left[H(t-t_z)\right] \over
H (t-t_z)^2} + \dots \; , \label{eq:asympbdot} \\
{\widehat F}(t) & = & 1 - {t_F \over t - t_z} + \dots \; , \label{eq:asympF}
\end{eqnarray}
where the higher corrections are down by inverse powers of $H (t-t_z)$,
possibly offset by logarithms of same. The parameters $H t_z$, $\alpha$ and 
$H t_F$ are determined by fitting the curves for each of the four runs. Their 
numerical values are reported in Table~3. Points to note are the close 
agreement between $N_{\rm pert}$ and $H t_z$, and the near constancy of 
$\alpha$ over three decades of variation in $\epsilon$. The first fact means 
that the number of e-foldings for inflation to end is well predicted by 
perturbation theory; the second fact means that the transition to radiation 
domination is almost independent of the scale of inflation. That the transition
is also quite rapid is illustrated by Figure~3. Note that it is {\it extremely} 
rapid with respect to the evolving time constant provided by $H_{\rm eff}(t)$ 
because the corrections fall off with the time constant $H \gg H_{\rm eff}(t)$.

\begin{table}

\vbox{\tabskip=0pt \offinterlineskip
\def\tablerule{\noalign{\hrule}}
\halign to450pt {\strut#& \vrule#\tabskip=1em plus2em& \hfil#& \vrule#& 
\hfil#\hfil& \vrule#& \hfil#& \vrule#& \hfil#\hfil& \vrule#& \hfil#& \vrule#& 
\hfil#\hfil& \vrule#\tabskip=0pt\cr
\tablerule
\omit & height2pt & \omit && \omit && \omit && \omit && \omit && \omit &\cr
&&\omit\hidewidth $\epsilon$ \hidewidth &&\omit\hidewidth $M$ ({\rm GeV})
\hidewidth&& \omit\hidewidth $N_{\rm pert}$ \hidewidth&& \omit\hidewidth 
$H t_z$ \hidewidth&& \omit\hidewidth $\alpha$ \hidewidth&& \omit\hidewidth 
$H t_F$ \hidewidth&\cr
\omit & height2pt & \omit && \omit && \omit && \omit && \omit && \omit &\cr
\tablerule
\omit & height2pt & \omit && \omit && \omit && \omit && \omit && \omit &\cr
&& $10^{-3}$ && $1.7 \times 10^{18}$ && $37.4$ && $38.1$ && $.2085$ && $.04959$
&\cr
\omit & height2pt & \omit && \omit && \omit && \omit && \omit && \omit &\cr
\tablerule
\omit & height2pt & \omit && \omit && \omit && \omit && \omit && \omit &\cr
&& $10^{-4}$ && $9.6 \times 10^{17}$ && $173.6$ && $174.9$ && $.2347$ && 
$.01208$ &\cr
\omit & height2pt & \omit && \omit && \omit && \omit && \omit && \omit &\cr
\tablerule
\omit & height2pt & \omit && \omit && \omit && \omit && \omit && \omit &\cr
&& $10^{-5}$ && $5.3 \times 10^{17}$ && $805.8$ && $808.7$ && $.2425$ && 
$.002680$ &\cr
\omit & height2pt & \omit && \omit && \omit && \omit && \omit && \omit &\cr
\tablerule
\omit & height2pt & \omit && \omit && \omit && \omit && \omit && \omit &\cr
&& $10^{-6}$ && $3.0 \times 10^{17}$ && $3740.$ && $3748.$ && $.2447$ && 
$.0005837$ &\cr
\omit & height2pt & \omit && \omit && \omit && \omit && \omit && \omit &\cr
\tablerule}}

\caption{Parameters from the various runs.}

\end{table}

\begin{figure}

\centerline{\psfig{figure=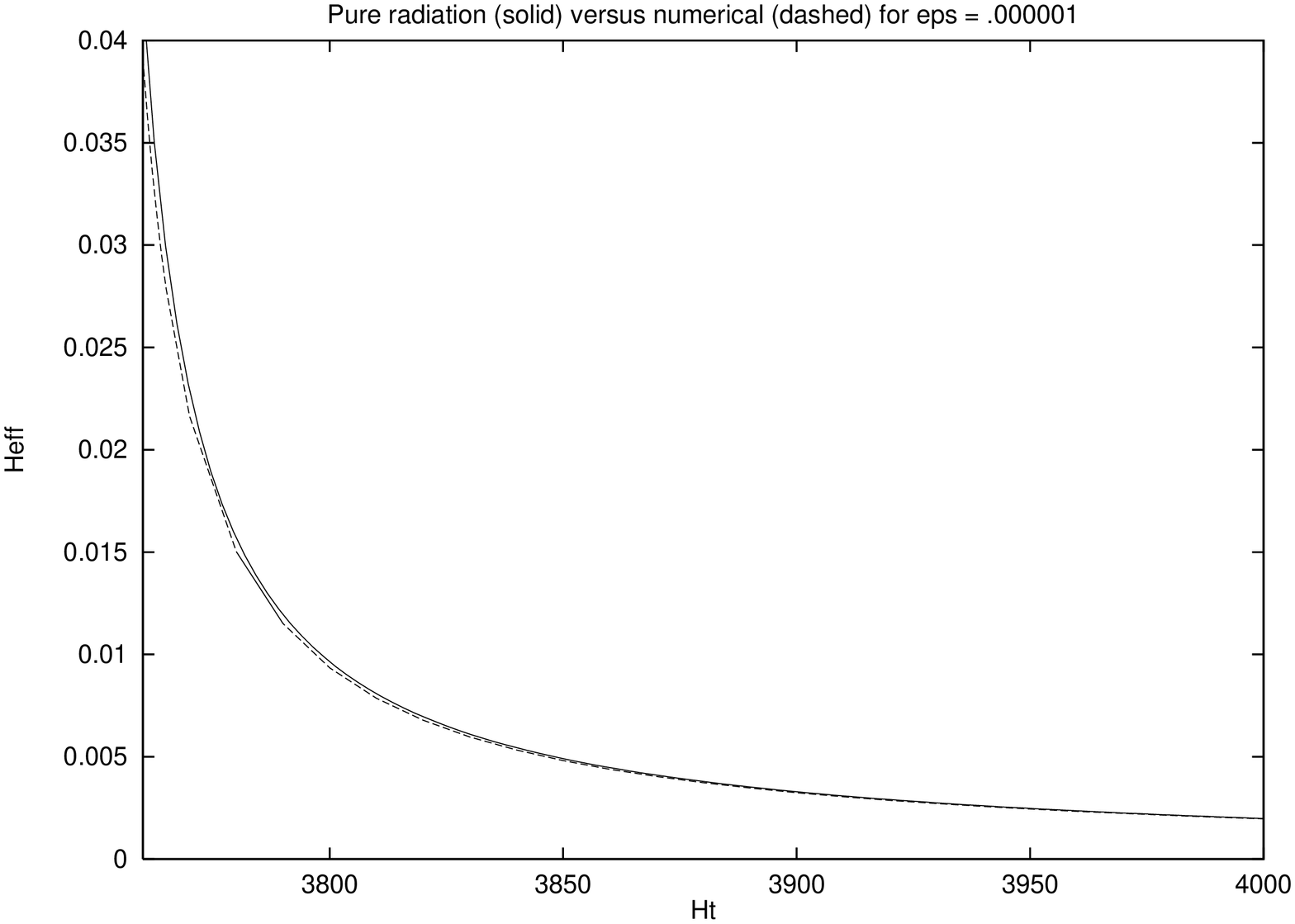,height=7cm,bbllx=0bp
,bblly=0bp,bburx=596bp,bbury=843bp,rheight=5.5cm,rwidth=10.5cm,angle=-90}
\hskip -1.5cm
\psfig{figure=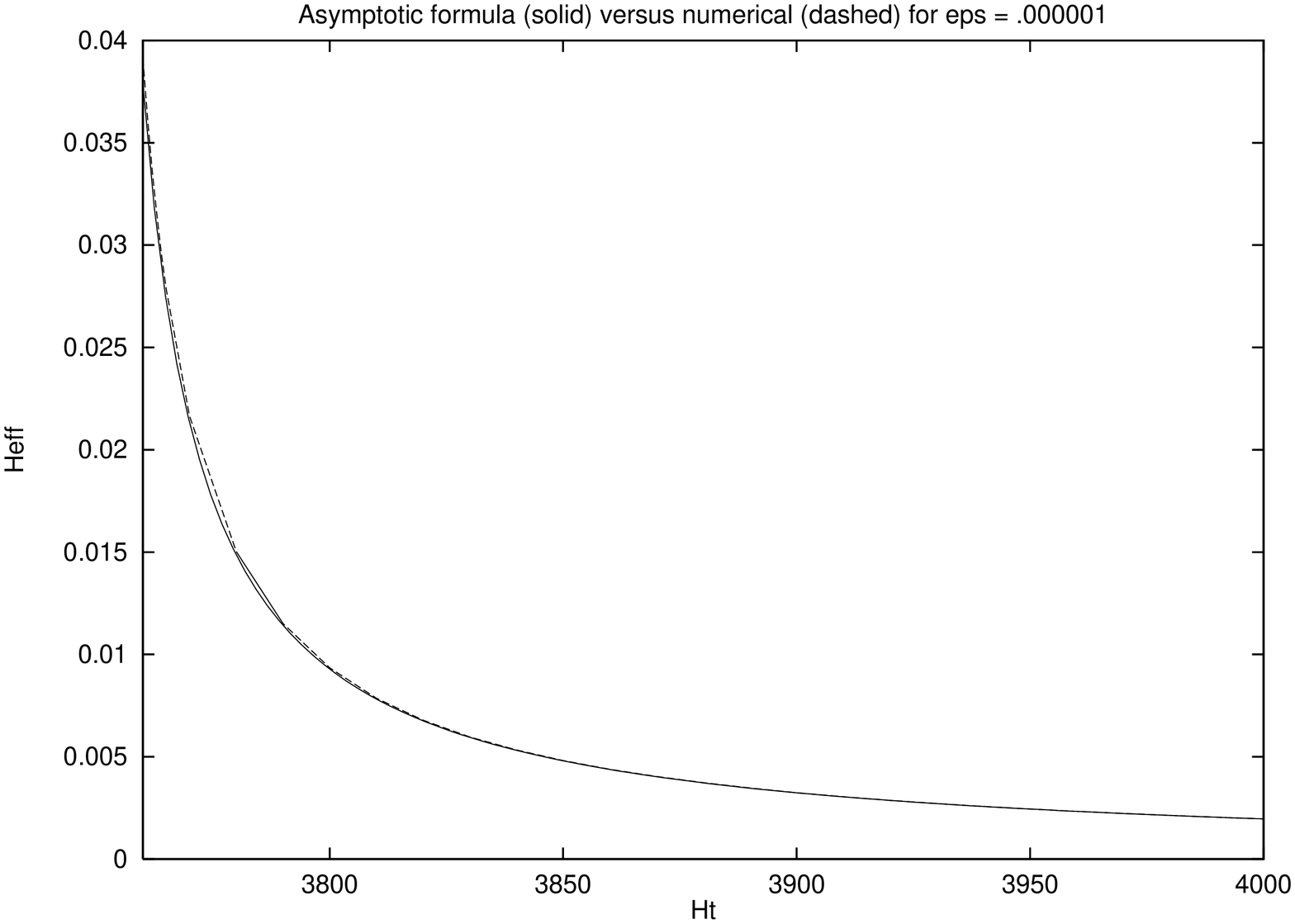,height=7cm,bbllx=0bp
,bblly=0bp,bburx=596bp,bbury=843bp,rheight=5.5cm,rwidth=10.5cm,angle=-90}}

\vskip 1cm

\caption{Asymptotic approach to radition domination for $\epsilon = 10^{-6}$.
Lefthand graph shows pure radiation (solid) versus the numerical result 
(dashed). The righthand graph show the asymptotic formula (solid) versus the 
numerical result (dashed).}

\end{figure}

Once square root expansion is accepted one can actually derive the asymptotic
series from the evolution equation. To see this, substitute the asymptotic
forms:
\begin{eqnarray}
\dot{b}(t) & = & {1 \over 2 (t - t_z)} + {\delta \dot{b}}(t) \; , 
\label{eq:asbdot} \\
{\widehat F}[b](t) & = & 1 - {\delta F}(t) \; , \label{eq:asF}
\end{eqnarray}
into the evolution equation (\ref{eq:Fform}). Taking the square root gives a 
differential equation for ${\delta F}(t)$:
\begin{equation}
{d \over dt} \ln({\delta F}) = - {1 \over t - t_z} + (t - t_z) {\delta 
\ddot{b}}(t) + \dots
\end{equation}
Integration gives the constant $t_F$ in (\ref{eq:asympF}). To get 
(\ref{eq:asympbdot}) we differentiate expression (\ref{eq:dFhat/dt}):
\begin{equation}
{d^2 {\widehat F} \over st^2} = -3 \dot{b} {d {\widehat F} \over dt} + 86 
\epsilon^2 \left(\ddot{b} + 2 \dot{b}^2\right) B^2[b](t) \; . \label{eq:d2F}
\end{equation}
Now substitute (\ref{eq:asympF}) and the leading term in ({\ref{eq:asympbdot})
to obtain the following leading order result:
\begin{equation}
{d^2 {\widehat F} \over dt^2} + 3 \dot{b} {d {\widehat F} \over dt} = - {t_F
\over 2 (t - t_F)^3} + \dots \; .
\end{equation}
This must be equal to $86 \epsilon^2 (\ddot{b} + 2 \dot{b}^2) B^2[b]$. The time
dependence must come from the term $\ddot{b}(t) + 2 \dot{b}^2(t)$ since 
$B[b](t)$ is dominated by what went on during inflation:
\begin{eqnarray}
B[b](t) & = & B[b](t_z) + \dots \; , \\
& = & H t_z + \dots \; .
\end{eqnarray}
The asymptotic form ({\ref{eq:asbdot}) gives:
\begin{equation}
\ddot{b}(t) + 2 \dot{b}^2(t) = {\delta \ddot{b}}(t) + {2 {\delta \dot{b}}(t)
\over t - t_z} + \dots \; .
\end{equation}
Substituting everything in (\ref{eq:d2F}) gives a differential equation for
${\delta b}(t)$:
\begin{equation}
{t_F \over 2 (t-t_z)^3} + \dots = 86 \epsilon^2 (H t_z)^2 \left({\delta 
\ddot{b}}(t) + {2 {\delta \dot{b}}(t) \over t - t_z}\right) + \dots \; .
\end{equation}
The solution is the second term in (\ref{eq:asympbdot}) with the relation:
\begin{equation}
H t_F \approx 172 \epsilon^2 (H t_z)^2 \alpha \; . \label{eq:relation}
\end{equation}

The independently fitted parameters of Table~3 obey (\ref{eq:relation}) to a
few percent, with agreement better for smaller values of $\epsilon$. The slight
discrepancy is probably due to the approximation $B[b](t) \approx B[b_{\rm 
class}](t_z) = H t_z$, which is an overestimate. If we assume that $\alpha$ is
almost constant then $H t_z \approx N_{\rm pert} \sim \epsilon^{-2/3}$ implies
that $H t_F$ varies as the two thirds power of $\epsilon$. A consequence is 
that the post-inflation value of ${\widehat F}[g](t)$ must be very close to one 
for low scale inflation.

It is perhaps significant that the effective Hubble constant $\dot{b}(t)$ 
approaches the expansion rate for a radiation dominated universe {\it from 
below}. Although matter is negligible during inflation, it cannot be ignored
afterwards. We must therefore expect that matter radiation is produced during
the transition. The asymptotic expansion rate means that this matter radiation 
is progressively enhanced with respect to the purely quantum gravitational 
stress energy we have been discussing.

\section{The scalar potential}

The purpose of this section is to reconstruct the scalar potential for the 
model we have just evolved. We begin by reviewing the general technique, then
we obtain analytic expressions during the perturbative and late time regimes.
The section closes with an explicit numerical reconstruction over the full
period of evolution.

Recall from Section 2 that the scalar potential $V(\phi)$ does not enter the
evolution equation and is therefore not required to determine $b(t)$. We 
instead reconstruct the potential from $b(t)$ by imposing stress energy 
conservation. The procedure is first to get the potential as a function of 
time:
\begin{equation}
{\widehat V}(t) = {1 \over 8 \pi G} \left(\ddot{b}(t) + 3 \dot{b}^2(t) -
3 H^2\right) \; . \label{eq:thepot}
\end{equation}
One then inverts the relation:
\begin{equation}
{\widehat \phi}(t) = -{1 \over \sqrt{8 \pi G}} \ln\left(1 - {\widehat F}[b](t)
\right) \; , \label{eq:thescalar}
\end{equation}
to find time as function of ${\widehat \phi}$. Substituting $t(\phi)$ in
(\ref{eq:thepot}) gives $V(\phi)$.

This procedure can be carried out analytically during the perturbative regime
where we can find explicit expressions for (\ref{eq:thepot}) and 
(\ref{eq:thescalar}):
\begin{eqnarray}
{\widetilde V}(t) & = & - {\Lambda \over 8 \pi G} { 3 \over N_{\rm pert}} 
{(Ht/N_{\rm pert})^2 \over 1 - (Ht/N_{\rm pert})^3} \left(1 - {1 \over 4 N_{\rm
pert}} {(Ht/N_{\rm pert})^2 \over 1 - (Ht/N_{\rm pert})^3}\right) 
\; , \;\;\;\;\;\;\; \\
{\widetilde \phi}(t) & = & - {1 \over \sqrt{8 \pi G}} \ln\left(1 - (Ht/N_{\rm
pert})^3\right) \; .
\end{eqnarray}
(Recall that $N_{\rm pert}$ is our perturbative estimate (\ref{eq:Npert}) for
the number of e-foldings of inflation.) Solving for the time as a function of 
the scalar:
\begin{equation}
{Ht \over N_{\rm pert}} = \left(1 - e^{-\sqrt{8 \pi G} \: {\widetilde \phi}}
\right)^{\frac13} \; ,
\end{equation}
gives the following relation for the potential during the perturbative regime:
\begin{eqnarray}
\lefteqn{V_{\rm pert}(\phi) =} \nonumber \\
& &  -{\Lambda \over 8 \pi G} {3 e^{\sqrt{8 \pi G} \phi} \over N_{\rm pert}} 
\left(1 - e^{-\sqrt{8 \pi G} \phi}\right)^{\frac23} \left[1 - {e^{\sqrt{8 \pi 
G} \phi} \over 4 N_{\rm pert}} \left(1 - e^{-\sqrt{8 \pi G} \phi}\right)^{
\frac23}\right] \; . \;\;\;\;\;\; 
\end{eqnarray}
This expression should be valid for $0 \leq \phi \ltwid 
\ln(N_{\rm pert})/\sqrt{8 \pi G}$. Note that it is the same for all models 
since it follows from the known perturbative results.

The other regime where explicit expressions can be obtained is that of late 
times. For the model of Section 6 we have:
\begin{eqnarray}
{\widehat V}(t) & = & - {\Lambda \over 8 \pi G} \left(1 - {1 \over 12 H^2
(t - t_z)^2} + \dots \right) \; , \\
{\widehat \phi}(t) & = & - {1 \over \sqrt{8 \pi G}} \ln\left({t_F \over t -t_z}
+ \dots \right) \; .
\end{eqnarray}
During this period the time can be expressed in terms of the scalar as:
\begin{equation}
t = t_z + t_F e^{\sqrt{8 \pi G} \: {\widehat \phi}} + \dots \; ,
\end{equation}
which gives the following result for the late time potential:
\begin{equation}
V_{\rm late}(\phi) = -{\Lambda \over 8 \pi G} \left[1 - \frac1{12} 
\left({e^{-\sqrt{8 \pi G} \phi} \over H t_F}\right)^2 + \dots \right] \; .
\end{equation}
This expression should be valid for $\phi \gtwid -\ln(H t_F)/\sqrt{8 \pi G}$.
Using (\ref{eq:relation}) we see that this is about $\phi \gtwid \ln(N_{\rm
pert})/\sqrt{8 \pi G}$, so most of the range is covered by the two asymptotic
expressions.

\begin{figure}

\centerline{\psfig{figure=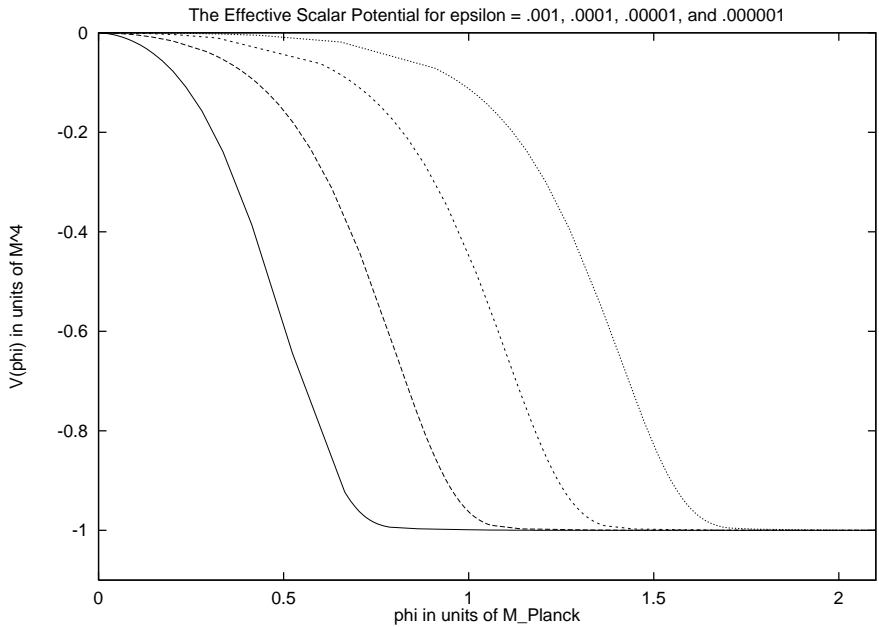,height=10cm,bbllx=0bp
,bblly=0bp,bburx=596bp,bbury=843bp,rheight=5.5cm,rwidth=12.7cm,angle=-90}}

\vskip 4cm

\caption{$V(\phi)$ versus $\phi$ for $\epsilon = 10^{-3}$ (leftmost), 
$10^{-4}$, $10^{-5}$ and $10^{-6}$ (rightmost).}

\end{figure}

Full coverage can be obtained by computing the values of $\phi$ and $V(\phi)$ 
at each step in the numerical evolution, and then plotting the resulting curve. 
The discretized formulae are:
\begin{eqnarray}
{\phi_i \over M_P} & = & -{1 \over \sqrt{8 \pi}} \ln(1 - F_i) \; , \\
{V_i \over M^4} & = & 1 - \left({{\Delta b}_i \over {\Delta \tau}}\right)^2
- \frac13 {{\Delta^2 b}_i \over {\Delta \tau}^2} \; .
\end{eqnarray}
The curves for each of the four runs are plotted in Figure~4. Note that they
confirm the results of our asymptotic expansions, including the fact that the
transition from inflation occurs at $\phi \sim \ln(N_{\rm pert})/\sqrt{8 \pi 
G}$.

\section{Discussion}

We have developed a compelling physical picture for a particular second order 
back-reaction that quantum gravity has on inflation. The first order effect is 
that long wavelength virtual gravitons can be ripped apart by the superluminal 
expansion of spacetime. This is the phenomenon of superadiabatic amplification, 
first studied by Grishchuk \cite{grish1}. This first order effect is not
secular. As more and more graviton pairs are injected into the inflating 
universe, the growth of their energy is cancelled by the expansion of the 
3-volume to produce a small, constant energy density (and pressure $p = -\rho$)
of magnitude about $H^4$. The secular effect comes at the next order from the
gravitational interaction between the receding virtual gravitons. As graviton 
pairs are pulled apart their long range gravitational potentials fill the 
intervening space, and these potentials remain to add with those of new pairs 
even after the old pairs have been redshifted into insignificance. The second
order effect is suppressed by the small dimensionless coupling constant $G 
\Lambda \ltwid 10^{-11}$, but it is cummulative. And it slows inflation because 
gravity is attractive. 

An explicit two loop computation has already confirmed that the quantum 
gravitational back-reaction slows inflation by an amount which eventually 
becomes non-perturbatively large \cite{tw2}. The question is, what happens 
next? We have argued in this paper that the question can be answered by 
inferring and then numerically evolving the most cosmologically significant
terms in the effective field equations. Section 2 proved that the important
part of the quantum gravitationally induced stress tensor is that of an 
effective scalar field $\phi[g]$ which is itself a non-local functional of the 
metric. We also showed that a model is completely specified by $\phi[g]$, since
the potential $V(\phi)$ does not enter the evolution equation, and we obtained
an explicit expression (\ref{eq:pertF}) for the scalar during the perturbative 
regime. 

In Sections 3 and 4 we showed that general considerations and the physics of 
screening very largely constrain the scalar. It must have the form:
\begin{equation}
\phi[g] = - {1 \over \sqrt{8 \pi G}} \ln\left(1 - F[g]\right) \; ,
\end{equation}
where the functional $F[g]$ consists of a retarded scalar Green's function 
$\Box^{-1}$ acting on a source composed of two more factors of $\Box^{-1}$ 
acting in some order on from three to five Ricci scalars. We did not determine 
whether the two inner Green's functions act in series or in parallel, nor did 
we fix how the various Ricci scalars are located with respect to them. The
resulting combinatorics yields 73 possibilities, many of which were tabulated 
in Section 5. We also identified two procedures for actually {\it deriving} the
scalar from the result of perturbative  computations of the same complexity as 
the one already done.

In Section 6 we selected one of the candidate models:
\begin{equation}
F[g] = - \frac{43}{48} \left({G \Lambda \over 3 \pi}\right)^2 {1 \over \Box}
\left( R \left({1 \over \Box} R\right)^2 \right) \; ,
\end{equation}
and numerically evolved it through the end of inflation into the regime of 
asymptotically late times. We found that inflation ends over a period of about 
five e-foldings, following which the universe asymptotically approaches the 
square root expansion of a radiation dominated universe:
\begin{equation}
\dot{b}(t) \longrightarrow {1 \over 2 ( t - t_z)} \; .
\end{equation}
Almost identical results were found for the series analog model:
\begin{equation}
F[g] = - \frac{43}{24} \left({G \Lambda \over 3 \pi}\right)^2 {1 \over \Box}
\left( R \left({1 \over \Box} R \left({1 \over \Box} R\right)\right)\right)\; ,
\end{equation}
although we did not report them. We were able to confirm the asymptotic forms 
by deriving analytic expressions from the non-linear and non-local evolution 
equation. An interesting consequence of this analysis is that the approach is 
from below, implying that any matter radiation produced in the transition would
be relatively enhanced by the slower redshift.

The asymptotic analysis can be used to categorize models by the number of 
factors of $R$ which lie immediately to the right of the outer $\Box^{-1}$. 
Suppose there are $f$ outer factors of $R$. We can expose them by taking 
derivatives:
\begin{equation}
-{\widehat {\Box}} {\widehat F} = {d^2 {\widehat F} \over dt^2} + 3 \dot{b}
{d {\widehat F} \over dt} \; . \label{eq:exposed}
\end{equation}
The two internal retarded Green's functions will be effectively constant, 
dominated by what went on during inflation, so the time dependence must come 
from the $f$ outer factors of ${\widehat R}$. Suppose that the asymptotic 
expansion rate is a general power law $p > 0$:
\begin{equation}
\dot{b}(t) \longrightarrow {p \over t - t_z} \; .
\end{equation} 
The Ricci scalar goes to:
\begin{equation}
{\widehat R}(t) = 6 \ddot{b}(t) + 12 \dot{b}^2(t) \longrightarrow {2 p^2 - p 
\over (t - t_z)^2} \; ,
\end{equation}
so the lefthand side of (\ref{eq:exposed}) goes like $(t -t_z)^{-2f}$, unless
$p = 1/2$. To find the righthand side, consider the evolution equation: 
\begin{equation}
\ddot{b}(t) = -\frac12 \left({d{\widehat F}/dt \over 1 - {\widehat F}}\right)^2
\; ,
\end{equation}
It follows that ${\widehat F}[b](t)$ must have the form:
\begin{equation}
{\widehat F}[b](t) \longrightarrow 1 - \left({t_F \over t - t_z}
\right)^{\sqrt{2p}} \; .
\end{equation}
Hence the righthand side of (\ref{eq:exposed}) goes like $(t - t_z)^{-2 -
\sqrt{2p}}$, and we must have $2 + \sqrt{2p} = 2 f$, unless $p = 1/2$.

It follows that the case of no outer Ricci scalars ($f=0$) is not consistent 
with stable evolution. What happens for these models is that ${\widehat 
F}[b](t)$ actually reaches one and $\ddot{b}(t)$ goes to minus infinity. For 
$f=1$ we get $p=0$, which is also not consistent. However, taking account of 
next order terms in the asymptotic expansion of $\dot{b}$ gives $p=1/2$. The 
other cases are all consistent: for $f=2$ we get $p=2$; $f=3$ gives $p=8$; 
$f=4$ results in $p= 18$; and $f=5$ produces $p=32$. Since the actual model is 
likely to be a linear combination of the tabulated candidates, and since the 
term with the lowest value of $f$ dominates the asymptotic behavior, we 
conclude that the approach to a radiation dominated universe is generic for 
models which include at least one $f=1$ term while avoiding any $f=0$ 
terms.\footnote{The $f=0$ terms on Table~1 are 16a-21a and 10b-12b; the $f=1$ 
terms are 11a-15a and 7b-9b. On Table~2 the $f=0$ terms are 7c-10c and 5d-6d; 
the $f=1$ terms are 4c-6c and 3d-4d.}

Of course fixing the asymptotic time dependence does not provide a physical
interpretation for what happens after the end of inflation. For example, the
stress energy of our model is not likely to consist of gravitons, in spite of
the fact that it approaches the equation of state of radiation. There is simply
no way to have created them. The energy density of gravitons produced by
superadiabatic amplification is a small constant for as long as it can be 
reliably tracked using perturbation theory --- which is almost to the end of
inflation. What terminates inflation is the buildup of gravitational 
interaction stress energy and it must be this which adds with the stress tensor
of the cosmological constant to produce a residual obeying the equation of 
state of radiation.

The effective scalar $\phi[g]$ poses a similar interpretational dilemma. It is 
certainly not a fundamental particle but one might plausibly interpret 
screening in terms of the formation of a scalar bound state on cosmological 
scales. The ``scalars'' would be the virtual gravitons pairs which are ripped
apart by inflation. Although they are separated to cosmological distances it is
their binding energy which eventually arrests inflation. Since the pairs tend
to annihilate on sub-horizon scales there are no new massless quanta to 
embarrass phenomenology. One very attractive feature of this interpretation is 
that we can compute the spectrum of density perturbations using the standard
formalism, just as if the scalar was fundamental.

Finally, there is the question of how to couple matter and what effect doing so
will have, both on the geometry and on reheating. Gravitons dominate screening 
through their unique combination of masslessness without conformal invariance, 
however, ordinary matter becomes important at the end of inflation. The 
simplest assumption would that the matter and the gravitational stress tensors
are separately conserved except for a brief period at the end of inflation when
some of the gravitational stress energy excites matter degrees of freedom. It
remains to see if there can be sufficient reheating without the phase of
coherent oscillations that characterizes scalar based inflation 
\cite{Linde,kt}.

\begin{center}
{\bf Acknowledgements}
\end{center}

We have profited from conversations with R. Abramo, R. H. Bran\-den\-berg\-er, 
V. M. Mukhanov, H. B. Nielsen and P. Sikivie. We thank the University of Crete 
for its hospitality during the execution of this project. This work was 
partially supported by DOE contract DE-FG02-97ER41029, by NSF grant 94092715, 
by EEC grant 961206, by NATO Collaborative Research Grant 971166 and by the 
Institute for Fundamental Theory.

\end{document}